\documentclass[preprint,prd,aps,showpacs,showkeys,nofootinbib]{revtex4}
\usepackage{graphicx}
\usepackage{dcolumn}
\usepackage{bm}
\usepackage{diagbox}
\usepackage{color}
\usepackage[dvipsnames]{xcolor}
\usepackage{amssymb}
\usepackage{makecell}
\definecolor{light-gray}{gray}{0.88}
\definecolor{dark-gray}{gray}{0.40}
\begin{document}

\title{Lepton flavor violating decays $l_j\rightarrow{l_i\gamma}$ in the $U(1)_X$SSM model within the Mass Insertion Approximation}
\author{Tong-Tong Wang$^{1,2}$\footnote{wtt961018@163.com}, Shu-Min Zhao$^{1,2}$\footnote{zhaosm@hbu.edu.cn}, Jian-Fei Zhang$^{1,2}$\footnote{zjf09@hbu.edu.cn}, Xing-Xing Dong$^{1,2}$\footnote{dongxx@hbu.edu.cn}, Tai-Fu Feng$^{1,2,3}$\footnote{fengtf@hbu.edu.cn}}

\affiliation{$^1$ Department of Physics, Hebei University, Baoding 071002, China}
\affiliation{$^2$ Key Laboratory of High-precision Computation and Application of Quantum Field Theory of Hebei Province, Baoding 071002, China}
\affiliation{$^3$ Department of Physics, Chongqing University, Chongqing 401331, China}
\date{\today}

\begin{abstract}

 Three singlet new Higgs superfields and right-handed neutrinos are added to MSSM to obtain $U(1)_X$SSM model. Its local
 gauge group is $SU(3)_C\times SU(2)_L \times U(1)_Y \times U(1)_X$.
In the framework of $U(1)_X$SSM, we study muon anomalous magnetic moment and lepton flavor violating decays $l_j\rightarrow{l_i\gamma}(j=2,3;i=1,2)$ within the Mass Insertion Approximation(MIA). Through the MIA method, we can find the parameters that directly affect the analytical result of the lepton flavor violating decays $l_j\rightarrow{l_i\gamma}$, which make our work more convenient. We want to provide a set of simple analytic formulas for the form factors and the
associated effective vertices, that may be very useful for future phenomenological
studies of the lepton flavor violating decays.
  According to the accuracy of the numerical results which the influence of different sensitive parameters, we come to the conclusion that the non-diagonal elements which correspond to the generations of the initial lepton and final lepton
are main sensitive parameters and lepton flavor violation(LFV) sources. This work can provide a clear signal of new physics(NP).

\end{abstract}


\maketitle

\section{introduction}
 Lepton has unitary matrix similar to  Cabbibo-Kobayashi-Maskawa(CKM) mixed matrix. The breaking theory of electric weak symmetry and neutrino oscillation experiment show that lepton flavor violation (LFV) exists both theoretically and experimentally\cite{1}. The standard model(SM) is already a mature theory. However, the lepton number is conserved in the SM, so there is no LFV process in the SM\cite{2}.
 Through research, it is necessary to expand the SM.
Any sign of LFV can be regarded as evidence of the existence of new physics(NP)\cite{02}.

Physicists have extended SM and obtained a large number of extended models, among which the minimum supersymmetric standard model (MSSM) is the most concerned model. However, it is gradually found that MSSM also has problems: $\mu$ problem\cite{3} and zero mass neutrinos\cite{UU1}. To solve these problems, we pay attention to the U(1) expansions of MSSM.
 We extend the MSSM with $U(1)_X$ gauge group, whose symmetry group is $SU(3)_C\times SU(2)_L \times U(1)_Y\times U(1)_X$\cite{Sarah1,Sarah2,Sarah3}. It adds three Higgs singlet superfields and right-handed neutrino superfields beyond MSSM\cite{MSSM}. There are five neutral CP-even Higgs component fields in the model, which come from two Higgs doublets and three Higgs singlets respectively. Therefore, the mass mixing matrix is $5\times5$, and the 125.1 GeV Higgs particle\cite{sm1} corresponds to the lightest mass eigenstate.

To improve the corrections to LFV processes of $l_j\rightarrow{l_i\gamma}$, people discuss different SM extended models\cite{220725}, for example minimal R-symmetric supersymmetric standard model\cite{9}, MSSM extension with gauged baryon and lepton numbers
\cite{21}, SM extension with a hidden $U(1)_X$ gauged symmetry\cite{new1} and lepton numbers and supersymmetric low-scale seesaw models\cite{new2}. It is worth noting that in our previous work, we have studied lepton flavor violating decays $l_j\rightarrow{l_i\gamma}$ in the $U(1)_X$SSM model\cite{20}. The above works and most of the research on LFV are studied with the mass eigenstate method. Using this method to find sensitive parameters is often not intuitive and clear enough, which depends on the mass eigenstates of the particles and rotation matrixes. It will lead us to pay too much attention to many unimportant parameters. Now we use a novel calculation method called as Mass Insertion Approximation(MIA)\cite{04,07,05,06}, which uses the electroweak interaction eigenstate and treats perturbatively the mass insertions changing slepton flavor.
By means of mass insertions inside the propagators of the electroweak interaction sleptons eigenstates, at the analytical level, we can find many parameters that have direct impact on LFV. It is worth noting that these parameters are considered between all possible flavor blends among SUSY partner of leptons, in which their particular origin has no assumption and is independent of the model\cite{07}. In addition, the MIA method has been applied to other works related to LFV, including the $h, H, A \rightarrow{\tau\mu}$ decays induced from SUSY loops\cite{07}, effective lepton flavor violating $H\ell_i\ell_j$ vertex from right-handed
neutrinos\cite{06}, one-loop effective LFV $Zl_kl_m$ vertex from heavy neutrinos\cite{05} and so on.
This method provides very simple and intuitive analytical formula, and is also clear about the changes of the main parameters affecting lepton taste destruction, which provides a new idea for other work of LFV in the future.

In the process of LFV, because the mass of $\tau$ lepton is much greater than $\mu$ and $e$, there are more LFV decay channels\cite{08}.
The decay processes of $l_j\rightarrow{l_i\gamma}$ are the most interesting.
This work is to study the LFV of the $l_j\rightarrow{l_i\gamma}$ processes under the $U(1)_X$SSM model. The effects of different reasonable parameter spaces on the branching ratio Br($l_j\rightarrow{l_i\gamma}$) are compared. The latest upper limits on the LFV branching ratio of $\mu\rightarrow{e\gamma}$,~$\tau\rightarrow{\mu\gamma}$ and $\tau\rightarrow{e\gamma}$ at 90\% confidence level (C.L.)\cite{10} are
\begin{eqnarray}
Br(\mu\rightarrow{e\gamma})<4.2\times10^{-13},~~Br(\tau\rightarrow{\mu\gamma})<4.4\times10^{-8},~~Br(\tau\rightarrow{e\gamma})<3.3\times10^{-8}.
\end{eqnarray}

The paper is organized as follows. In Sec.II, we mainly introduce the $U(1)_X$SSM including its superpotential and the general soft breaking terms. In Sec.III, we give analytic expressions for muon anomalous magnetic moment and the branching ratios of $l_j\rightarrow{l_i\gamma}$ decays in the $U(1)_X$SSM. In Sec.IV, we give the numerical analysis, and the summary is given in Sec.V.

\section{the $U(1)_X$SSM}

$U(1)_X$SSM is the U(1) extension of MSSM, whose local gauge group is  $SU(3)_C\otimes
SU(2)_L \otimes U(1)_Y\otimes U(1)_X$\cite{04,UU1,UU3}. On the basis of MSSM, $U(1)_X$SSM has new superfields such as three Higgs
singlets $\hat{\eta},~\hat{\bar{\eta}},~\hat{S}$ and right-handed neutrinos $\hat{\nu}_i$.
 Through the seesaw mechanism, light
neutrinos obtain tiny masses at the tree level.
The neutral CP-even parts of
$H_u,~ H_d,~\eta,~\bar{\eta}$ and $S$ mix together and form a $5\times5 $ mass squared matrix, whose
lightest mass eigenvalue corresponds to the lightest CP-even Higgs. The particle content and charge assignments for $U(1)_X$SSM can be found in our previous work\cite{UU1}.
 To get 125.1 GeV Higgs mass\cite{LCTHiggs1,LCTHiggs2},
the loop corrections should be taken into account. The sneutrinos are disparted into CP-even sneutrinos and CP-odd sneutrinos,
and their mass squared matrixes are both extended to $6\times6$.

In $U(1)_X$SSM, the concrete form of the superpotential is:
\begin{eqnarray}
&&W=l_W\hat{S}+\mu\hat{H}_u\hat{H}_d+M_S\hat{S}\hat{S}-Y_d\hat{d}\hat{Q}\hat{H}_d-Y_e\hat{e}\hat{L}\hat{H}_d+\lambda_H\hat{S}\hat{H}_u\hat{H}_d
\nonumber\\&&~~~~~~+\lambda_C\hat{S}\hat{\eta}\hat{\bar{\eta}}+\frac{\kappa}{3}\hat{S}\hat{S}\hat{S}+Y_u\hat{u}\hat{Q}\hat{H}_u+Y_X\hat{\nu}\hat{\bar{\eta}}\hat{\nu}
+Y_\nu\hat{\nu}\hat{L}\hat{H}_u.
\end{eqnarray}
We collect the explicit forms of two Higgs doublets and three Higgs singlets here
\begin{eqnarray}
&&H_{u}=\left(\begin{array}{c}H_{u}^+\\{1\over\sqrt{2}}\Big(v_{u}+H_{u}^0+iP_{u}^0\Big)\end{array}\right),
~~~~~~
H_{d}=\left(\begin{array}{c}{1\over\sqrt{2}}\Big(v_{d}+H_{d}^0+iP_{d}^0\Big)\\H_{d}^-\end{array}\right),
\nonumber\\
&&\eta={1\over\sqrt{2}}\Big(v_{\eta}+\phi_{\eta}^0+iP_{\eta}^0\Big),~~~~~~~~~~~~~~~
\bar{\eta}={1\over\sqrt{2}}\Big(v_{\bar{\eta}}+\phi_{\bar{\eta}}^0+iP_{\bar{\eta}}^0\Big),\nonumber\\&&
\hspace{4.0cm}S={1\over\sqrt{2}}\Big(v_{S}+\phi_{S}^0+iP_{S}^0\Big).
\end{eqnarray}
The vacuum expectation values(VEVs) of the Higgs superfields $H_u$, $H_d$, $\eta$, $\bar{\eta}$ and $S$
are denoted by $v_u,~v_d,~v_\eta$,~ $v_{\bar\eta}$ and $v_S$ respectively. Two angles are defined as
$\tan\beta=v_u/v_d$ and $\tan\beta_\eta=v_{\bar{\eta}}/v_{\eta}$.

The soft SUSY breaking terms of this model are shown as
\begin{eqnarray}
&&\mathcal{L}_{soft}=\mathcal{L}_{soft}^{MSSM}-B_SS^2-L_SS-\frac{T_\kappa}{3}S^3-T_{\lambda_C}S\eta\bar{\eta}
+\epsilon_{ij}T_{\lambda_H}SH_d^iH_u^j\nonumber\\&&\hspace{1.5cm}
-T_X^{IJ}\bar{\eta}\tilde{\nu}_R^{*I}\tilde{\nu}_R^{*J}
+\epsilon_{ij}T^{IJ}_{\nu}H_u^i\tilde{\nu}_R^{I*}\tilde{L}_j^J
-m_{\eta}^2|\eta|^2-m_{\bar{\eta}}^2|\bar{\eta}|^2-m_S^2S^2\nonumber\\&&\hspace{1.5cm}
-(m_{\tilde{\nu}_R}^2)^{IJ}\tilde{\nu}_R^{I*}\tilde{\nu}_R^{J}
-\frac{1}{2}\Big(M_X\lambda^2_{\tilde{X}}+2M_{BB^\prime}\lambda_{\tilde{B}}\lambda_{\tilde{X}}\Big)+h.c~~.
\end{eqnarray}

 We have proven that $U(1)_X$SSM is  anomaly free in our previous work\cite{UU3}.
 Two Abelian groups $U(1)_Y$ and $U(1)_X$ produce a new effect called as the gauge kinetic mixing in the $U(1)_X$SSM, which is MSSM never before.

In general, the covariant derivatives of $U(1)_X$SSM can be written as \cite{UMSSM5,B-L1,B-L2,gaugemass}
\begin{eqnarray}
&&D_\mu=\partial_\mu-i\left(\begin{array}{cc}Y,&X\end{array}\right)
\left(\begin{array}{cc}g_{Y},&g{'}_{{YX}}\\g{'}_{{XY}},&g{'}_{{X}}\end{array}\right)
\left(\begin{array}{c}A_{\mu}^{\prime Y} \\ A_{\mu}^{\prime X}\end{array}\right)\;,
\label{gauge1}
\end{eqnarray}
 with $A_{\mu}^{\prime Y}$ and $A^{\prime X}_\mu$ representing the gauge fields of $U(1)_Y$ and $U(1)_X$ respectively.

Under the condition that the two Abelian gauge groups are unbroken, we use the rotation
matrix $R$ \cite{UMSSM5,B-L2,gaugemass} to perform a change of the basis
\begin{eqnarray}
&&D_\mu=\partial_\mu-i\left(\begin{array}{cc}Y^Y,&Y^X\end{array}\right)
\left(\begin{array}{cc}g_{Y},&g{'}_{{YX}}\\g{'}_{{XY}},&g{'}_{{X}}\end{array}\right)R^TR
\left(\begin{array}{c}A_{\mu}^{\prime Y} \\ A_{\mu}^{\prime X}\end{array}\right)\;,
\end{eqnarray}
with the redefinitions
\begin{eqnarray}
&&\left(\begin{array}{cc}g_{Y},&g{'}_{{YX}}\\g{'}_{{XY}},&g{'}_{{X}}\end{array}\right)
R^T=\left(\begin{array}{cc}g_{1},&g_{{YX}}\\0,&g_{{X}}\end{array}\right)~~~~\text{and}~~~~~
R\left(\begin{array}{c}A_{\mu}^{\prime Y} \\ A_{\mu}^{\prime X}\end{array}\right)
=\left(\begin{array}{c}A_{\mu}^{Y} \\ A_{\mu}^{X}\end{array}\right)\;.
\end{eqnarray}
Then the covariant derivatives of $U(1)_X$SSM are changed as
\begin{eqnarray}
&&D_\mu=\partial_\mu-i\left(\begin{array}{cc}Y^Y,&Y^X\end{array}\right)
\left(\begin{array}{cc}g_{1},&g_{{YX}}\\0,&g_{{X}}\end{array}\right)
\left(\begin{array}{c}A_{\mu}^{Y} \\ A_{\mu}^{X}\end{array}\right)\;.
\end{eqnarray}

At the tree level, three neutral gauge bosons $A^{X}_\mu,~A^Y_\mu$ and $V^3_\mu$ mix together, whose mass matrix
is shown in the basis $(A^Y_\mu, V^3_\mu, A^{X}_\mu)$\cite{04}
\begin{eqnarray}
&&\left(\begin{array}{*{20}{c}}
\frac{1}{8}g_{1}^2 v^2 &~~~ -\frac{1}{8}g_{1}g_{2} v^2 & ~~~\frac{1}{8}g_{1}(g_{{YX}}+g_{X}) v^2 \\
-\frac{1}{8}g_{1}g_{2} v^2 &~~~ \frac{1}{8}g_{2}^2 v^2 & ~~~~-\frac{1}{8}g_{2}g_{{YX}} v^2\\
\frac{1}{8}g_{1}(g_{{YX}}+g_{X}) v^2 &~~~ -\frac{1}{8}g_{2}(g_{{YX}}+g_{X}) v^2 &~~~~ \frac{1}{8}(g_{{YX}}+g_{X})^2 v^2+\frac{1}{8}g_{{X}}^2 \xi^2
\end{array}\right),\label{gauge matrix}
\end{eqnarray}
with $v^2=v_u^2+v_d^2$ and $\xi^2=v_\eta^2+v_{\bar{\eta}}^2$.

 We use two mixing angles $\theta_{W}$ and $\theta_{W}'$ to get mass eigenvalues of the matrix in Eq.(\ref{gauge matrix}).
 $\theta_{W}$ is the Weinberg angle and the new mixing angle $\theta_{W}'$ is
 defined from the following formula

\begin{eqnarray}
\sin^2\theta_{W}'=\frac{1}{2}-\frac{[(g_{{YX}}+g_{X})^2-g_{1}^2-g_{2}^2]v^2+
4g_{X}^2\xi^2}{2\sqrt{[(g_{{YX}}+g_{X})^2+g_{1}^2+g_{2}^2]^2v^4+8g_{X}^2[(g_{{YX}}+g_{X})^2-g_{1}^2-g_{2}^2]v^2\xi^2+16g_{X}^4\xi^4}}.
\end{eqnarray}

It appears in the couplings involving $Z$ and $Z^{\prime}$.
The exact eigenvalues of Eq.(\ref{gauge matrix}) are deduced \cite{04}
\begin{eqnarray}
&&m_\gamma^2=0,\nonumber\\
&&m_{Z,{Z^{'}}}^2=\frac{1}{8}\Big([g_{1}^2+g_2^2+(g_{{YX}}+g_{X})^2]v^2+4g_{X}^2\xi^2 \nonumber\\
&&\hspace{1.1cm}\mp\sqrt{[g_{1}^2+g_{2}^2+(g_{{YX}}+g_{X})^2]^2v^4+8[(g_{{YX}}+g_{X})^2-g_{1}^2-
g_{2}^2]g_{X}^2v^2\xi^2+16g_{X}^4\xi^4}\Big).
\end{eqnarray}

The used mass matrixes can be found in the work\cite{UU1,20}.
Here, we show some needed couplings in this model.
We deduce the vertexes of $\bar{l}_i-\chi_j^--\tilde{\nu}^R_k(\tilde{\nu}^I_k)$
\begin{eqnarray}
&&\mathcal{L}_{\bar{l}\chi^-\tilde{\nu}^R}=\frac{1}{\sqrt{2}}\bar{l}_i\Big\{\tilde{\nu}^R_LY_l^iP_L\tilde{H}^{-}_1
-g_2\tilde{\nu}^R_LP_R\tilde{W}^{-}\Big\},\nonumber\\
&&\mathcal{L}_{\bar{l}\chi^-\tilde{\nu}^I}=\frac{i}{\sqrt{2}}\bar{l}_i\Big\{\tilde{\nu}^I_LY_l^iP_L\tilde{H}^{-}_1
-g_2\tilde{\nu}^I_LP_R\tilde{W}^{-}\Big\}.
\end{eqnarray}

We deduce the vertex couplings of neutralino-lepton-slepton
\begin{eqnarray}
&&\mathcal{L}_{\bar{\chi}^0l\tilde{L}}=\Big\{\Big(\frac{1}{\sqrt{2}}(g_1\lambda_{\tilde{B}}+g_2\tilde{W}^0+g_{YX}\lambda_{\tilde{X}})\tilde{L}^L
-\tilde{H}^0_dY^j_l\tilde{L}^R\Big)P_L\nonumber\\&&\hspace{1.6cm}
-\Big[\frac{1}{\sqrt{2}}\Big(2g_1\lambda_{\tilde{B}}+(2g_{YX}+g_X)\lambda_{\tilde{X}}\Big)\tilde{L}^R+\tilde{H}^0_dY_{l}^j\tilde{L}^L\Big]P_R\Big\}l_j.
\end{eqnarray}

\section {formulation}

In this section, we study the LFV of the $l_j\rightarrow{l_i\gamma}~(j=2,3;~i=1,2)$ and muon anomalous magnetic moment under the $U(1)_X$SSM model\cite{21} with the MIA. The simplified form is discussed.

\subsection{ Using MIA to calculate $l_j\rightarrow{l_i\gamma}$ in $U(1)_X$SSM model}
If the external lepton is on shell, the amplitude of $l_j\rightarrow{l_i\gamma}$ is
\begin{eqnarray}
\mathcal{M}=e\varepsilon^\mu\bar{u}_i(p+q)[q^2\gamma_\mu(C^L_1P_L+C^R_1P_R)+m_{l_j}i\sigma_{\mu\nu}q^\nu(C^L_2P_L+C^R_2P_R)]u_j(p),{\label{NI}}
\end{eqnarray}
 where $p$ is the injecting lepton momentum, $q$ is the photon momentum, and $m_{l_j}$ is the mass of the $j$th generation charged lepton. $\bar{u}_i(p)$ and $u_j(p)$ are the wave functions for the external leptons. The final Wilson coefficients $C^L_1,~C^R_1,~C^L_2,~C^R_2$ are obtained from the sum of these diagrams' amplitudes.
\begin{figure}[h]
\setlength{\unitlength}{5.0mm}
\centering
\includegraphics[width=5.0in]{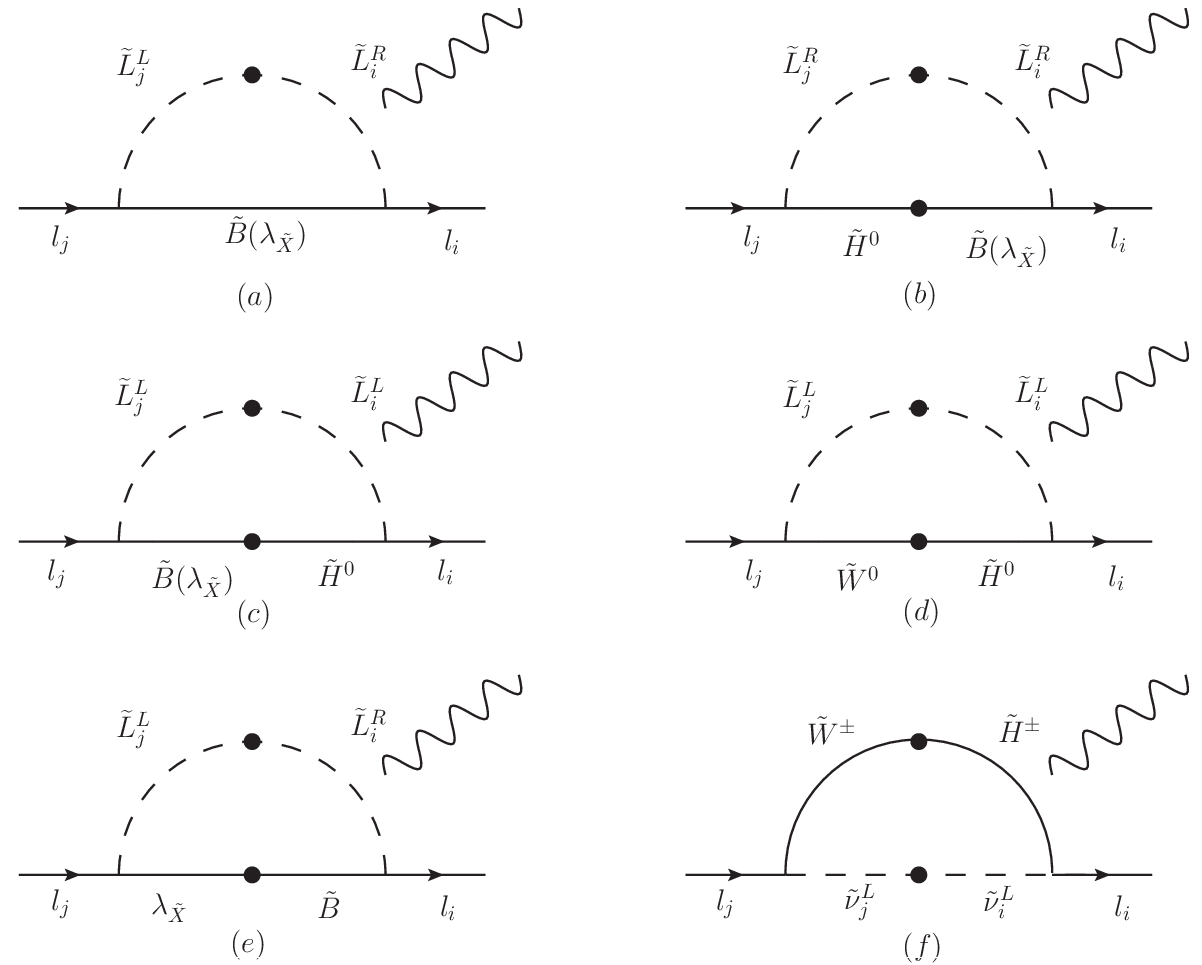}
\caption{  Feynman diagrams for $l_j\rightarrow{l_i\gamma}$ in the MIA.}\label{N1}
\end{figure}

The Feynman diagrams of $l_j\rightarrow{l_i\gamma}$ under the $U(1)_X$SSM model are obtained by MIA\cite{09} in Fig.\ref{N1}.  The sneutrinos are disparted
into CP-even sneutrinos $\tilde{\nu}^R$ and CP-odd sneutrinos $\tilde{\nu}^I$. After our analysis, in Fig.\ref{N1}(f) since right-handed sneutrinos are strongly depressed by $Y_{\nu}$, the situation of right-handed sneutrinos here is neglected.
In other words, there are only two cases of left-handed CP-even sneutrinos $\tilde{\nu}^R_L$ and left-handed  CP-odd sneutrinos $\tilde{\nu}^I_L$ in Fig.\ref{N1}(f).
In order to more directly express the influencing factors of LFV of $l_j\rightarrow{l_i\gamma}$, we use $C^f_2=C^{Lf}_2=C^{Rf}_2(f=1\cdots6)$ to express the one-loop corrections by MIA.

1. The one-loop contributions from $\tilde{B}(\lambda_{\tilde{X}})$-$\tilde{L}^L_j$-$\tilde{L}^R_i$.
\begin{eqnarray}
&&C_2^1(\tilde{L}^L_j,\tilde{L}^R_i,\tilde{B})=\frac{-1}{2m_{l_j}\Lambda^3}\Delta^{LR}_{ij}g_1^2\sqrt{x_1}[I_1(x_{\tilde{L}^L_j},x_1)+I_1(x_{\tilde{L}^R_i},x_1)\nonumber\\
&&\hspace{3.0cm}-2I_2(x_{\tilde{L}^L_j},x_1)-I_2(x_{\tilde{L}^R_i},x_1)],\label{MIABLR}
\\&&C_2^1(\tilde{L}^L_j,\tilde{L}^R_i, \lambda_{\tilde{X}})
=\frac{-1}{2m_{l_j}\Lambda^3}\Delta^{LR}_{ij}(g_{YX}^2+\frac{1}{2}g_{YX}g_X)\sqrt{x_{\lambda_{\tilde{X}}}}[I_1(x_{\tilde{L}^L_j},x_{\lambda_{\tilde{X}}})+I_1(x_{\tilde{L}^R_i},x_{\lambda_{\tilde{X}}})\nonumber\\
&&\hspace{3.0cm}-2I_2(x_{\tilde{L}^L_j},x_{\lambda_{\tilde{X}}})-I_2(x_{\tilde{L}^R_i},x_{\lambda_{\tilde{X}}})],\label{MIAXLR}
\end{eqnarray}
here, $m$ is the particle mass, with $x=\frac{m^2}{\Lambda^2}$. The functions $I_1(x,y)$ and $I_2(x,y)$ are
\begin{eqnarray}
&&I_1(x,y)=
   \frac{1}{32\pi^2}\Big\{\frac{1}{x(x-y)}-\frac{2+2\log x }{(x-y)^2 }+\frac{2x\log x-2y\log y }{(x-y)^3  }\Big\},
\\&&I_2(x,y)=
   \frac{1}{96\pi^2}\Big\{\frac{2 }{x(x-y)}-\frac{9+6\log x }{(x-y)^2 }+\frac{6x+12x\log x }{(x-y)^3  }
   \nonumber\\&&\hspace{1.8cm}-\frac{6x^2\log x-6y^2\log y}{(x-y)^4}\Big\}.
\end{eqnarray}

2. The one-loop contributions from $\tilde{B}(\lambda_{\tilde{X}})$-$\tilde{H}^0$-$\tilde{L}^R_j$-$\tilde{L}^R_i$.
\begin{eqnarray}
&&C_2^2(\tilde{L}^R_j,\tilde{L}^R_i,\tilde{B},\tilde{H}^0)
=\frac{1}{2\Lambda^4}g_1^2\tan\beta\sqrt{x_1x_{\mu^{\prime}_H}}\Delta^{RR}_{ij}[2I_4(x_{\tilde{L}^R_i},x_1,x_{\mu^{\prime}_H})\nonumber\\
&&\hspace{3.0cm}+3I_3(x_{\tilde{L}^R_j},x_1,x_{\mu^{\prime}_H})],\label{MIAHBR}
\\&&C_2^2(\tilde{L}^R_j, \tilde{L}^R_i, \lambda_{\tilde{X}}, \tilde{H}^0)
=\frac{1}{2\Lambda^4}\frac{1}{2}(2g_{YX}+g_X)(g_{YX}+g_X)\tan\beta\sqrt{x_{\lambda_{\tilde{X}}}x_{\mu^{\prime}_H}}\Delta^{RR}_{ij}\nonumber\\
&&\hspace{3.0cm}[2I_4(x_{\tilde{L}^R_i},x_{\lambda_{\tilde{X}}},x_{\mu^{\prime}_H})+3I_3(x_{\tilde{L}^R_j},x_{\lambda_{\tilde{X}}},x_{\mu^{\prime}_H})],\label{MIAHXR}
\end{eqnarray}
here $\mu_{H}^\prime=\frac{\lambda_H v_S}{\sqrt{2}}+\mu$
and $x_{\mu^{\prime}_H}=\frac{\mu_{H}^{\prime2}}{\Lambda^2}$.
The specific forms of $I_3(x,y,z)$ and $I_4(x,y,z)$ are
\begin{eqnarray}
&&I_3(x,y,z)=
   \frac{1}{32\pi^2}\Big\{\frac{6x+12x\log x }{(x-y) (x-z)^3 }+\frac{6x+12x\log x }{(x-y)^2 (x-z)^2 }+\frac{6x+12x\log x }{(x-y)^3 (x-z) }
   \nonumber\\&&\hspace{1.8cm}-\frac{9+6\log x}{(x-y) (x-z)^2}-\frac{9+6\log x}{(x-y)^2 (x-z)}+\frac{2}{x(x-y)(x-z)}
   \nonumber\\&&\hspace{1.8cm}+\frac{6x^2 \log x}{(x-y)(x-z)^4}+\frac{6x^2 \log x}{(x-y)^4(x-z)}-\frac{6x^2 \log x}{(x-y)^2(x-z)^3}
    \nonumber\\&&\hspace{1.8cm}-\frac{6x^2 \log x}{(x-y)^3(x-z)^2}+\frac{6y^2 \log y}{(x-y)^4(y-z)}-\frac{6z^2 \log z}{(y-z)(x-z)^4}\Big\},
\\&&I_4(x,y,z)=
   \frac{1}{32\pi^2}\Big\{-\frac{2+2\log x }{(x-y) (x-z)^2 }-\frac{2+2\log x }{(x-y)^2 (x-z) }+\frac{1 }{x(x-y) (x-z) }
   \nonumber\\&&\hspace{1.8cm}+\frac{2x\log x}{(x-y) (x-z)^3}+\frac{2x\log x}{(x-y)^2 (x-z)^2}+\frac{2x\log x}{(x-y)^3(x-z)}
    \nonumber\\&&\hspace{1.8cm}-\frac{2y \log y}{(x-y)^3(y-z)}+\frac{2z \log z}{(x-z)^3(y-z)}\Big\}.
\end{eqnarray}

3. The one-loop contributions from $\tilde{B}(\lambda_{\tilde{X}})$-$\tilde{H}^0$-$\tilde{L}^L_j$-$\tilde{L}^L_i$.
\begin{eqnarray}
&&C_2^3(\tilde{L}^L_j,\tilde{L}^L_i,\tilde{H}^0,\tilde{B})
=\frac{-m_{l_i}}{4m_{l_j}\Lambda^4}g_1^2\tan\beta\sqrt{x_1x_{\mu^{\prime}_H}}\Delta^{LL}_{ij}[2I_4(x_{\tilde{L}^L_i},x_1,x_{\mu^{\prime}_H})\nonumber\\
&&\hspace{3.0cm}+3I_3(x_{\tilde{L}^L_j},x_1,x_{\mu^{\prime}_H})],\label{MIABHL}
\\&&C_2^3(\tilde{L}^L_j,\tilde{L}^L_i,\tilde{H}^0,\lambda_{\tilde{X}})
=\frac{-m_{l_i}}{4m_{l_j}\Lambda^4}g_{YX}(g_{YX}+g_X)\tan\beta\sqrt{x_{\lambda_{\tilde{X}}}x_{\mu^{\prime}_H}}\Delta^{LL}_{ij}\nonumber\\
&&\hspace{3.0cm}[2I_4(x_{\tilde{L}^L_i},x_{\lambda_{\tilde{X}}},x_{\mu^{\prime}_H})+3I_3(x_{\tilde{L}^L_j},x_{\lambda_{\tilde{X}}},x_{\mu^{\prime}_H})].\label{MIAXHL}
\end{eqnarray}

4. The one-loop contributions from $\tilde{W}^0$-$\tilde{H}^0$-$\tilde{L}^L_j$-$\tilde{L}^L_i$.
\begin{eqnarray}
&&C_2^4(\tilde{L}^L_j,\tilde{L}^L_i, \tilde{H}^0, \tilde{W}^0)
=\frac{m_{l_i}}{4m_{l_j}\Lambda^4}g_2^2\tan\beta\sqrt{x_2x_{\mu^{\prime}_H}}\Delta^{LL}_{ij}\nonumber\\
&&\hspace{3.0cm}[2I_4(x_{\tilde{L}^L_i},x_2,x_{\mu^{\prime}_H})+3I_3(x_{\tilde{L}^L_j},x_2,x_{\mu^{\prime}_H})].\label{MIAWHL}
\end{eqnarray}

5. The one-loop contributions from $\tilde{B}-\lambda_{\tilde{X}}-\tilde{L}^L_j-\tilde{L}^R_i$.
\begin{eqnarray}
&&C_2^5(\tilde{L}^L_j,\tilde{L}^R_i, \tilde{B}, \lambda_{\tilde{X}})
=\frac{-1}{2m_{l_j}\Lambda^3}\Delta^{LR}_{ij}g_1g_{YX}\sqrt{x_{BB^\prime}x_1x_{\lambda_{\tilde{X}}}}[I_4(x_{\tilde{L}^L_j},x_1,x_{\lambda_{\tilde{X}}})\nonumber\\
&&\hspace{3.0cm}+I_4(x_{\tilde{L}^R_i},x_1,x_{\lambda_{\tilde{X}}})+I_5(x_{\tilde{L}^L_j},x_1,x_{\lambda_{\tilde{X}}})+2I_5(x_{\tilde{L}^R_i},x_1,x_{\lambda_{\tilde{X}}})]\nonumber\\
&&\hspace{3.0cm}+\frac{1}{2m_{l_j}\Lambda^3}\Delta^{LR}_{ij}g_1g_{YX}\sqrt{x_{BB^\prime}}[I_6(x_{\tilde{L}^L_j},x_1,x_{\lambda_{\tilde{X}}})\nonumber\\
&&\hspace{3.0cm}+I_6(x_{\tilde{L}^R_i},x_1,x_{\lambda_{\tilde{X}}})+I_7(x_{\tilde{L}^L_j},x_1,x_{\lambda_{\tilde{X}}})+2I_7(x_{\tilde{L}^R_i},x_1,x_{\lambda_{\tilde{X}}})].\label{MIAXBLR}
\end{eqnarray}

We show the one-loop functions $I_5(x,y,z)$ and $I_6(x,y,z)$ in the following form
\begin{eqnarray}
&&I_5(x,y,z)=\frac{-1}{32\pi^2}\Big\{\frac{3+2\log x }{(x-y)(x-z)}-\frac{2x+4x\log x}{(x-y)(x-z)^2}-\frac{2x+4x\log x}{(x-y)^2(x-z)}
 \nonumber\\&&\hspace{1.8cm}+\frac{2x^2\log x}{(x-y)(x-z)^3}+\frac{2x^2\log x}{(x-y)^2(x-z)^2}+\frac{2x^2\log x}{(x-y)^3(x-z)}
 \nonumber\\&&\hspace{1.8cm}-\frac{2y^2\log y}{(x-y)^3(y-z)}+\frac{2z^2\log z}{(x-z)^3(y-z)} \Big\},
\\&&I_6(x,y,z)=\frac{1}{96\pi^2}\Big\{\frac{6x^2 (3x^2+y^2+z^2+yz-3xy-3xz)(1+3\log x) }{(x-y)^3(x-z)^3}
\nonumber\\&&\hspace{1.8cm}-\frac{(6x^2-3xy-3xz)(5+6\log x)}{(x-y)^2(x-z)^2}+\frac{6y^3\log y-6x^3\log x}{(x-y)^4(y-z)}
\nonumber\\&&\hspace{1.8cm}+\frac{11+6\log x}{(x-y)(x-z)}+\frac{6x^3\log x-6z^3\log z}{(x-z)^4(y-z)}\Big\}.
\end{eqnarray}

 6. The one-loop contributions from chargino and left-handed CP-even(odd) sneutrino.
\begin{eqnarray}
&&C_2^6(\tilde{\nu}^I_{Lj},\tilde{\nu}^I_{Li}, \tilde{H}^{\pm}, \tilde{W}^{\pm})
=\frac{1}{2\Lambda^4}g_2^2\Delta^{LL}_{ij}\tan\beta\{(\sqrt{x_2x_{\mu^{\prime}_H}}+x_{\mu^{\prime}_H})I_8(x_{\mu^{\prime}_H},x_2,x_{\tilde{\nu}^I_{Li}})\nonumber\\
&&\hspace{3.0cm}+(\sqrt{x_2x_{\mu^{\prime}_H}}+x_2)I_8(x_2,x_{\mu^{\prime}_H},x_{\tilde{\nu}^I_{Lj}})\nonumber\\
&&\hspace{3.0cm}+\sqrt{x_2x_{\mu^{\prime}_H}}I_9(x_2,x_{\mu^{\prime}_H},x_{\tilde{\nu}^I_{Li}})-I_{10}(x_2,x_{\mu^{\prime}_H},x_{\tilde{\nu}^I_{Lj}})\},\label{MIARC}
\\&&C_2^6(\tilde{\nu}^R_{Lj},\tilde{\nu}^R_{Li}, \tilde{H}^{\pm}, \tilde{W}^{\pm})
=\frac{1}{2\Lambda^4}g_2^2\Delta^{LL}_{ij}\tan\beta\{(\sqrt{x_2x_{\mu^{\prime}_H}}+x_{\mu^{\prime}_H})I_8(x_{\mu^{\prime}_H},x_2,x_{\tilde{\nu}^R_{Li}})\nonumber\\
&&\hspace{3.0cm}+(\sqrt{x_2x_{\mu^{\prime}_H}}+x_2)I_8(x_2,x_{\mu^{\prime}_H},x_{\tilde{\nu}^R_{Lj}})\nonumber\\
&&\hspace{3.0cm}+\sqrt{x_2x_{\mu^{\prime}_H}}I_9(x_2,x_{\mu^{\prime}_H},x_{\tilde{\nu}^R_{Li}})-I_{10}(x_2,x_{\mu^{\prime}_H},x_{\tilde{\nu}^R_{Lj}})\}.\label{MIAIC}
\end{eqnarray}

The one-loop functions $I_7(x,y,z)$,$I_8(x,y,z)$ and $I_9(x,y,z)$  read as
\begin{eqnarray}
&&I_7(x,y,z)=\frac{-1}{32\pi^2}\Big[\frac{8x\log x -4x}{(x-y) (x-z)^3}-\frac{2x+4x \log x}{(x-y)^2 (x-z)^2}+\frac{3+2\log x}{(x-y) (x-z)^2}
\nonumber\\&&\hspace{1.8cm}+\frac{2z+4z\log z}{ (x-z)^3 (y-z)}+\frac{4x^2\log x}{(x-y)^2(x-z)^3}-\frac{2x^2\log x}{(x-y)^3(x-z)^2}
   \nonumber\\&&\hspace{1.8cm}+\frac{2z^3\log z}{(x-z)^3(y-z)^2}-\frac{2y^2\log y}{(x-y)^3(y-z)^2}+\frac{6x^2\log x}{(x-y)(x-z)^4}
   \nonumber\\&&\hspace{1.8cm}+\frac{6z^2\log z}{(x-z)^4(y-z)}\Big],
\\&&I_8(x,y,z)=\frac{-1}{16\pi^2}\Big[\frac{x+2x\log x}{(x-y)^2 (x-z)^2}+\frac{y+2y\log y}{(x-y)^2(y-z)^2}+\frac{z+2z\log z}{(x-z)^2(y-z)^2}
\nonumber\\&&\hspace{1.8cm}-\frac{2x^2 \log x}{(x-y)^2 (x-z)^3}-\frac{2x^2\log x}{(x-y)^3(x-z)^2}-\frac{2y^2 \log y}{(x-y)^2(y-z)^3}
\nonumber\\&&\hspace{1.8cm}+\frac{2y^2\log y}{(x-y)^3(y-z)^2}+\frac{2z^2\log z}{(x-z)^2(y-z)^3}+\frac{2z^2\log z}{(x-z)^3(y-z)^2}\Big],
\\&&I_9(x,y,z)=\frac{1}{96\pi^2}\Big[\frac{x^2+3x^2\log x}{(x-y) (x-z)^2}+\frac{3z^2\log z-z^2}{(x-z)^2(y-z)}-\frac{2x^3 \log x}{(x-y) (x-z)^3}
\nonumber\\&&\hspace{1.8cm}-\frac{2z^3 \log z}{(x-z)^3(y-z)}-\frac{x^3\log x}{(x-y)^2(x-z)^2}+\frac{y^3\log y}{(x-y)^2(y-z)^2}
\nonumber\\&&\hspace{1.8cm}-\frac{z^3\log z}{(x-z)^2(y-z)^2}\Big].
\end{eqnarray}

From the above formulas, we can find that $C_2^f(f=1\cdots6)$ are mostly affected by $\tan\beta$ and $\Delta^{AB}_{ij}(A,B=L,R)$ and there is a positive correlation. $\Delta^{AB}_{ij}$ have the lepton flavor violating sources. It provides a reference for our subsequent work.
Finally, we get the final Wilson coefficient and decay width of $l_j\rightarrow{l_i\gamma}$,
\begin{eqnarray}
&&C_2=\sum^{i=1\cdots6}_iC_2^i,\nonumber\\
&&\Gamma(l_j\rightarrow{l_i\gamma})=\frac{e^2}{8\pi}m^5_{l_j}|C_2|^2.
\end{eqnarray}

The beanching ratio of $l_j\rightarrow{l_i\gamma}$ is
\begin{eqnarray}
Br(l_j\rightarrow{l_i\gamma})=\Gamma(l_j\rightarrow{l_i\gamma})/\Gamma_{l_j}.
\end{eqnarray}

\subsection{ Degenerate Result}

In order to more intuitively analyze the factors affecting lepton flavor violating processes $l_j\rightarrow{l_i\gamma}$,  we suppose that all the masses of the superparticles are almost degenerate.
 In other words, we give the one-loop results (chargino-sneutrino, neutralino-slepton) in the extreme case, where the masses for superparticles ($M_1,~M_2,~\mu_H^\prime,~m_{\tilde{L}_L}~
,m_{\tilde{L}_R},~M_{\lambda_{\tilde{X}}},~M_{BB^\prime} $) are equal to $M_{SUSY}$\cite{04}:
\[M_1=|M_2|=\mu_H^\prime=m_{\tilde{L}_L}
=m_{\tilde{L}_R}=M_{\lambda_{\tilde{X}}}=|M_{BB^\prime}|=M_{SUSY}.\]

The functions $I_i(i=1\cdots9)$ and $\Delta^{AB}_{ij}(A,B=L,R)$ are much simplified as
 \begin{eqnarray}
&&I_1(1,1)=\frac{-1}{96\pi^2},\hspace{2.2cm}I_2(1,1)=\frac{-1}{192\pi^2},\hspace{1.8cm}I_3(1,1,1)=\frac{-1}{480\pi^2},\nonumber\\
&&I_4(1,1,1)=\frac{1}{192\pi^2},\hspace{1.5cm}I_5(1,1,1)=\frac{1}{192\pi^2},\hspace{1.5cm}I_6(1,1,1)=\frac{-1}{320\pi^2},\nonumber\\
&&I_7(1,1,1)=\frac{-1}{480\pi^2},\hspace{1.5cm}I_8(1,1,1)=\frac{-1}{480\pi^2},\hspace{1.5cm}I_9(1,1,1)=\frac{1}{384\pi^2},
\\&&\Delta^{LR}_{ij}=m_{l_j}m_{\tilde{L}_L}\delta^{LR}_{ij},\hspace{1.8cm}\Delta^{LL}_{ij}=m^2_{\tilde{L}_L}\delta^{LL}_{ij},\hspace{2.2cm}\Delta^{RR}_{ij}=m^2_{\tilde{L}_R}\delta^{RR}_{ij}.
\end{eqnarray}

Then, we obtain the much simplified one-loop results of $C_2$
 \begin{eqnarray}
&&C_2=\frac{(2g_1^2\texttt{sign}[M_1\mu_H^{\prime}]+(2g_{YX}^2+3g_{YX}g_X+g_X^2)\texttt{sign}[M_{\lambda_{\tilde{X}}}\mu_H^{\prime}])\tan\beta\delta^{RR}_{ij}}{960\pi^2M^2_{SUSY}}\nonumber\\
&&+\frac{(-g_1^2\texttt{sign}[M_1\mu_H^{\prime}]-(g_{YX}^2+g_{YX}g_X)\texttt{sign}[M_{\lambda_{\tilde{X}}}\mu_H^{\prime}]+g_2^2\texttt{sign}[M_2\mu_H^{\prime}])m_{l_i}\tan\beta\delta^{LL}_{ij}}{960\pi^2M^2_{SUSY}m_{l_j}}\nonumber\\
&&+\frac{(-4g_2^2\texttt{sign}[M_2^2]-4g_2^2\texttt{sign}[\mu_{H}^{\prime2}]-12g_2^2\texttt{sign}[\mu_{H}^{\prime}M_2]+5g_2^2)\tan\beta\delta^{LL}_{ij}}{3840\pi^2M^2_{SUSY}}\nonumber\\
&&+\frac{1}{1920\pi^2M^2_{SUSY}}\times\{(5g_1^2\texttt{sign}[M_1]+5(g_{YX}^2+\frac{1}{2}g_{YX}g_X)\texttt{sign}[M_{\lambda_{\tilde{X}}}]\nonumber\\
&&-4g_1g_{YX}\texttt{sign}[M_{BB^{\prime}}M_1M_{\lambda_{\tilde{X}}}]+g_1g_{YX}\texttt{sign}[M_{BB^{\prime}}])\delta^{LR}_{ij}\}.
\end{eqnarray}

In the above formula, after simple approximation, we can find that in the formula of the second line, due to the existence of $\frac{m_{l_i}}{m_{l_j}}$, the result is 2-3 orders of magnitude smaller than other terms. Therefore, we will not consider the term with $\frac{m_{l_i}}{m_{l_j}}$ here. It can be found that $\texttt{sign}[M_1],\texttt{sign}[M_2],\texttt{sign}[M_{\lambda_{\tilde{X}}}],\texttt{sign}[\mu_H^{\prime}]$ and $\texttt{sign}[M_{BB^{\prime}}]$ have a certain impact on the correction of $C_2$.
According to  $1>g_X>g_{YX}>0$, we assume $\texttt{sign}[M_1]=\texttt{sign}[M_{\lambda_{\tilde{X}}}]=\texttt{sign}[\mu_H^{\prime}]=1$ and $\texttt{sign}[M_2]=\texttt{sign}[M_{BB^{\prime}}]=-1$, and get the larger value of $C_2$
\begin{eqnarray}
&&C_2=\frac{(5g_1^2+5(g_{YX}^2+\frac{1}{2}g_{YX}g_X)+3g_1g_{YX})\delta^{LR}_{ij}}{1920\pi^2M^2_{SUSY}}+\frac{3g_2^2\tan\beta\delta^{LL}_{ij}}{1280\pi^2M^2_{SUSY}}\nonumber\\
&&\hspace{0.8cm}+\frac{(2g_1^2+(2g_{YX}^2+3g_{YX}g_X+g_X^2))\tan\beta\delta^{RR}_{ij}}{960\pi^2M^2_{SUSY}}.\nonumber\\
\end{eqnarray}

 Due to the different orders of magnitude of branching ratios, we set $\tan\beta=9,~M_{SUSY}=1000$ GeV and discuss in two cases:

 1.$\mu\rightarrow e\gamma$

\begin{figure}[ht]
\setlength{\unitlength}{5mm}
\centering
\includegraphics[width=4.0in]{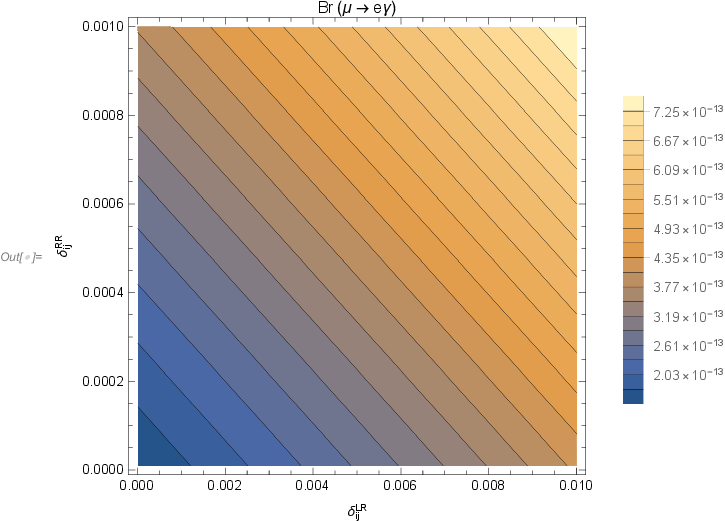}
\vspace{0.2cm}
\caption{Under the condition that $g_{YX}=0.2,~g_X=0.3$ and $\delta^{LL}_{ij}=1\times10^{-3}$, the effect of $\delta^{LR}_{ij}$ and $\delta^{RR}_{ij}$ on $Br(\mu\rightarrow e\gamma)$. The x-axis representing the range of $\delta^{LR}_{ij}$ is from $1\times10^{-5}$ to $1\times10^{-2}$, and the y-axis represents $1\times10^{-5}<\delta^{RR}_{ij}<1\times10^{-3}$. The rightmost icon is the color corresponding to the value of $Br(\mu\rightarrow e\gamma)$.}{\label {N2}}
\end{figure}

 We take $g_{YX},g_X,\delta^{LR}_{ij},\delta^{LL}_{ij}$ and $\delta^{RR}_{ij}$ as variables to study the effect on $Br(\mu\rightarrow e\gamma)$. In Fig.\ref{N2}, it can be found that both $\delta^{LR}_{ij}$ and $\delta^{RR}_{ij}$ have great influence on $Br(\mu\rightarrow e\gamma)$ and they are all increasing trend. The larger the values of $\delta^{LR}_{ij}$ and $\delta^{RR}_{ij}$, the easier it is to approach the upper limit of the experiment.

\begin{figure}[ht]
\setlength{\unitlength}{5mm}
\centering
\includegraphics[width=4.0in]{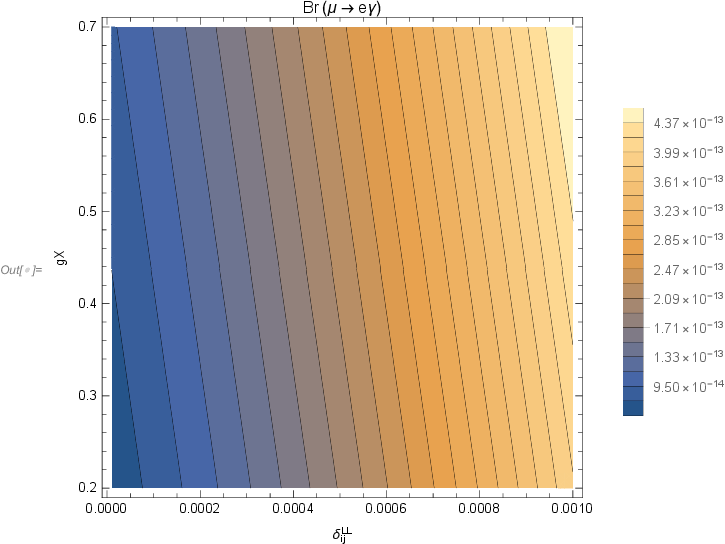}
\vspace{0.2cm}
\caption{Under the condition that $g_{YX}=0.2,~\delta^{RR}_{ij}=1\times10^{-6}$ and $\delta^{LR}_{ij}=1\times10^{-2}$, $\delta^{LL}_{ij}$ versus $g_X$ about $Br(\mu\rightarrow e\gamma)$. The abscissa is  $1\times10^{-5}<\delta^{LL}_{ij}<1\times10^{-3}$ and the ordinate represents $0.2<g_X<0.7$. The icon on the right shows the value of $Br(\mu\rightarrow e\gamma)$.}{\label {N3}}
\end{figure}

Similar to the above, in Fig.\ref{N3}, with the increase of $\delta^{LL}_{ij}$, the value of $Br(\mu\rightarrow e\gamma)$ gradually increases, and when $g_X$ increases, $Br(\mu\rightarrow e\gamma)$ also increases. When $\delta^{LL}_{ij}=9\times10^{-4}$ and $g_X=0.65$, $Br(\mu\rightarrow e\gamma)$ reaches the experimental upper limit. But in numerical terms, the effect of $\delta^{LL}_{ij}$ is greater than $g_X$.

 2.$\tau\rightarrow \mu(e)\gamma$

\begin{figure}[ht]
\setlength{\unitlength}{5mm}
\centering
\includegraphics[width=4.0in]{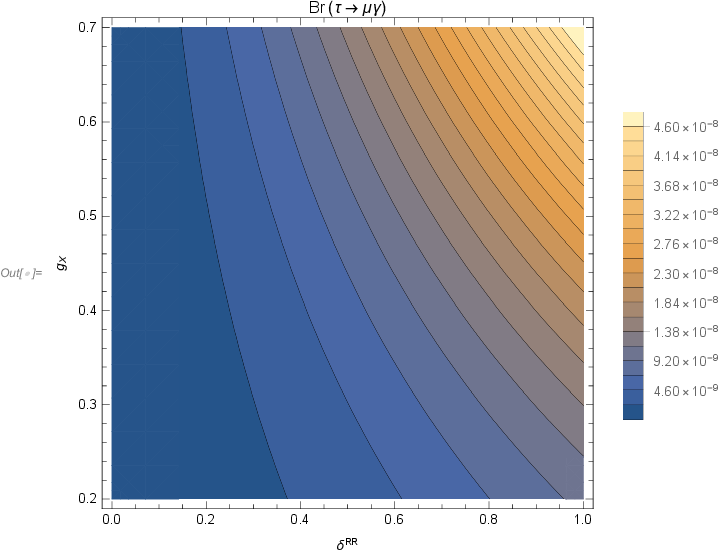}
\vspace{0.2cm}
\caption{Under the condition that $\delta^{LR}_{ij}=0.1,~g_{YX}=0.2$ and $\delta^{LL}_{ij}=0.1$, the effect of $\delta^{RR}_{ij}$ and $g_X$ on $Br(\tau\rightarrow\mu\gamma)$. The x-axis representing the range of $\delta^{RR}_{ij}$ is from $10^{-5}$ to 1, and the y-axis represents $0.2<g_X<0.7$. The rightmost icon is the color corresponding to the value of $Br(\tau\rightarrow\mu\gamma)$.}{\label {N4}}
\end{figure}

 Since the numerical results of $Br(\tau\rightarrow\mu\gamma)$ and $Br(\tau\rightarrow e\gamma)$ are close and have similar characteristic, we take $\tau\rightarrow\mu\gamma$ as an example. In Fig.\ref{N4}, when the values of $\delta^{RR}_{ij}$ and $g_X$ enlarger, the value of $Br(\tau\rightarrow\mu\gamma)$ also increases, which can well reach the experimental measured value.

\begin{figure}[ht]
\setlength{\unitlength}{5mm}
\centering
\includegraphics[width=4.0in]{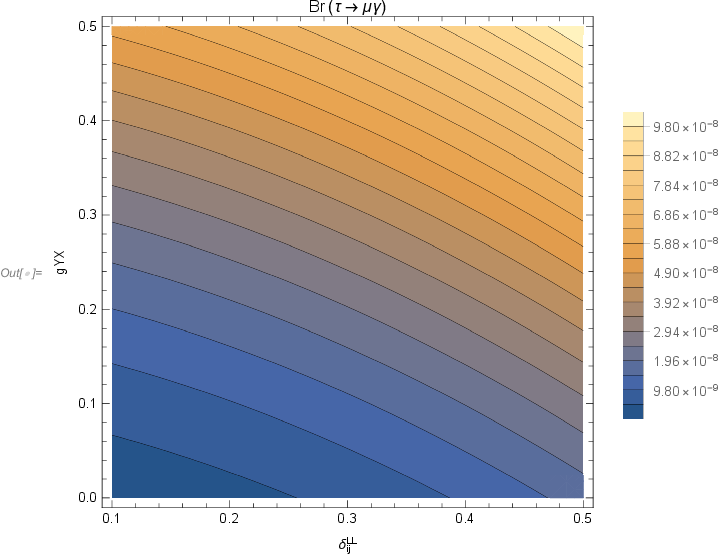}
\vspace{0.2cm}
\caption{Under the condition that $\delta^{RR}_{ij}=0.2,~g_X=0.3$ and $\delta^{LR}_{ij}=0.5$, $\delta^{LL}_{ij}$ versus $g_{YX}$ about $Br(\tau\rightarrow\mu\gamma)$. The abscissa is  $1\times10^{-5}<\delta^{LL}_{ij}<0.5$ and the ordinate represents $0.1<g_{YX}<0.5$. The icon on the right shows the value of $Br(\tau\rightarrow\mu\gamma)$.}{\label {N5}}
\end{figure}

In Fig.\ref{N5}, we analyze $\delta^{LL}_{ij}$ and $g_{YX}$ on the $Br(\tau\rightarrow\mu\gamma)$. The value of $Br(\tau\rightarrow\mu\gamma)$ also increases with the increasing $\delta^{LL}_{ij}$ and $g_{YX}$, but the effect from  $g_{YX}$ is greater than $\delta^{LL}_{ij}$. So the correction of $g_{YX}$ to $Br(\tau\rightarrow\mu\gamma)$ is greater than that of $\delta^{LL}_{ij}$.

All in all, we can find that $g_{YX},~g_X,~\delta^{LR}_{ij},~\delta^{LL}_{ij}$ and $\delta^{RR}_{ij}$  all have direct impact on the correction to $Br(\mu\rightarrow e\gamma),~Br(\tau\rightarrow\mu\gamma)$ and $Br(\tau\rightarrow e\gamma)$.

\subsection{Muon anomalous magnetic moment}
The one-loop corrections to muon anomalous magnetic moment are obtained with MIA.
Here, we show the one-loop contributions from chargino and CP-even(odd) sneutrino as\cite{04}
\begin{eqnarray}
&&a_\mu(\tilde{\nu}^R_L, \tilde{H}^{\pm}, \tilde{W}^{\pm})
=\frac{g_2^2}{2}
x_\mu\sqrt{x_2x_{\mu^{\prime}_H}}\tan\beta[2\mathcal{I}(x_{\mu^{\prime}_H},x_{\tilde{\nu}^R_L},x_2)
-\mathcal{J}(x_2,x_{\mu^{\prime}_H},x_{\tilde{\nu}^R_L})
\nonumber\\&&\hspace{3.8cm}+2\mathcal{I}(x_2,x_{\tilde{\nu}^R_L},x_{\mu^{\prime}_H})
-\mathcal{J}(x_{\mu^{\prime}_H},x_2,x_{\tilde{\nu}^R_L})]\label{MIARC},
\\&&a_\mu(\tilde{\nu}^I_L, \tilde{H}^{\pm}, \tilde{W}^{\pm})
=\frac{g_2^2}{2}
x_\mu\sqrt{x_2x_{\mu^{\prime}_H}}\tan\beta[2\mathcal{I}(x_{\mu^{\prime}_H},x_{\tilde{\nu}^I_L},x_2)
-\mathcal{J}(x_2,x_{\mu^{\prime}_H},x_{\tilde{\nu}^I_L})
\nonumber\\&&\hspace{3.8cm}+2\mathcal{I}(x_2,x_{\tilde{\nu}^I_L},x_{\mu^{\prime}_H})
-\mathcal{J}(x_{\mu^{\prime}_H},x_2,x_{\tilde{\nu}^I_L})]\label{MIAIC}.
\end{eqnarray}

The concrete forms of the one-loop functions $\mathcal{I}(x,y,z)$ and $\mathcal{J}(x,y,z)$ are
\begin{eqnarray}
&&\mathcal{J}(x,y,z)=\frac{1}{16\pi^2}\Big[\frac{x (x^2+x z-2 y
   z)\log x}{(x-y)^2 (x-z)^3}-\frac{y^2 \log
   y}{(x-y)^2 (y-z)^2}\nonumber\\&&\hspace{1.8cm}+\frac{z[x
   (z-2 y)+z^2] \log z}{(z-x)^3 (y-z)^2}-\frac{x
   (y-2 z)+y z}{(x-y) (x-z)^2 (y-z)}\Big].
\\&&\mathcal{I}(x,y,z)=\frac{1}{16\pi^2}\Big[\frac{1}{(x-z) (z-y)}+\frac{(z^2-x y)\log z}{(x-z)^2
   (y-z)^2}\nonumber\\&&\hspace{1.8cm}-\frac{x \log
   x}{(x-y) (x-z)^2}+\frac{y \log y}{(x-y)
   (y-z)^2}\Big].
\end{eqnarray}

The other one-loop contributions are obtained from $\tilde{B}(\lambda_{\tilde{X}})$-$\tilde{L}_L$-$\tilde{L}_R$,
$\tilde{B}(\lambda_{\tilde{X}})$-$\tilde{H}^0$-$\tilde{L}_R$, $\tilde{B}(\tilde{W}^0,\lambda_{\tilde{X}})$-$\tilde{H}^0$-$\tilde{L}_L$ and $\tilde{B}-\lambda_{\tilde{X}}-\tilde{L}_R-\tilde{L}_L$.
To save space in the text, we do not show their concrete forms here, which can be found in Ref.\cite{04}.
In our previous work\cite{04}, the one-loop contributions of muon anomalous magnetic moment in the degenerate form
are get with the supposition $M_1=M_2=\mu_H^\prime=m_{\tilde{L}_L}
=m_{\tilde{L}_R}=|M_{\lambda_{\tilde{X}}}|=|M_{BB^\prime}|=M_{SUSY}$
\begin{eqnarray}
&&a^{1L}_\mu\simeq \frac{1}{192\pi^2}\frac{m_\mu^2}{M_{SUSY}^2}\tan\beta(5g_2^2+g_1^2)\nonumber\\&&
+\frac{1}{960\pi^2}\frac{m_\mu^2}{M_{SUSY}^2}\tan\beta\Big[5(g_{YX}^2-g_{YX}g_X-g_X^2)\texttt{sign}[M_{\lambda_{\tilde{X}}}]
\nonumber\\&&+g_1(4g_{YX}+g_X)\texttt{sign}[M_{BB^\prime}]
\Big(1-4\texttt{sign}[M_{\lambda_{\tilde{X}}}]\Big)\Big].\label{amuS}
\end{eqnarray}

\section{numerical results}

In this section, we study the numerical results and consider the constraints from lepton flavor violating processes $l_j\rightarrow l_i\gamma$. In addition, we have considered the following conditions:
1. the lightest CP-even Higgs mass $m_{h^0}$=125.1 GeV\cite{su1,su2}.
2. The latest experimental results of the mass of the heavy vector boson $Z^\prime$ is $M_{Z^{\prime}}> 5.1$ TeV\cite{xin1}.
3. The limits for the masses of other particles beyond SM.
4. The bound on the ratio between $M_{Z^\prime}$ and its gauge coupling $g_X$ is $M_{Z^\prime}/g_X\geq6$ TeV at 99\% C.L.\cite{ZPG1,ZPG2}.
5. The constraint from LHC data, $\tan \beta_\eta<1.5$\cite{TanBP}.
6. The scalar lepton masses larger than 700 GeV and chargino masses larger than 1100 GeV\cite{A2021}.

Considering the above constraints in the front paragraph, we use the following parameters
\begin{eqnarray}
&&M_S =2.7 ~{\rm TeV},~
T_{\kappa} =1.6~ {\rm TeV}, ~
M_1 =1.2~{\rm TeV},~ M_2=M_{BL}=1~{\rm TeV},
~g_{YX}=0.2,\nonumber\\&&
\xi = 17~{\rm TeV},~Y_{X11} =Y_{X22} = Y_{X33} =1,~
g_X=0.3,~\kappa=1,~\lambda_C = -0.08,~v_S=4.3~{\rm TeV},
\nonumber\\&& M_{BB^\prime}=0.4~{\rm TeV},~
~T_{\lambda_H} = 0.3~{\rm TeV},~
M^2_{\tilde{L}11}= M^2_{\tilde{L}22} = M^2_{\tilde{L}33}=M^2_{\tilde{L}}= 0.5~{\rm TeV^2},~
 l_W = 4~{\rm TeV}^2,
 \nonumber\\&&\lambda_H = 0.1,~
T_{e11} =T_{e22} =T_{e33}= 5~{\rm TeV},~\tan\beta_\eta=0.8,~ B_{\mu} = B_S=1~{\rm TeV}^2,~\mu=0.5~{\rm TeV},\nonumber\\&&
T_{\lambda_C} = -0.1~{\rm TeV},~M^2_{\tilde{E}11}= M^2_{\tilde{E}22}= M^2_{\tilde{E}33}=M^2_{\tilde{E}}=3.6~{\rm TeV}^2.
\end{eqnarray}
To simplify the numerical research, we use the relations for the parameters and they vary in the following numerical analysis
\begin{eqnarray}
&&~M^2_{\tilde{L}12}=M^2_{\tilde{L}21},~M^2_{\tilde{L}13}=M^2_{\tilde{L}31},M^2_{\tilde{L}32}=M^2_{\tilde{L}23},\nonumber\\
&&~M^2_{\tilde{E}12}=M^2_{\tilde{E}21},~M^2_{\tilde{E}13}=M^2_{\tilde{E}31},~M^2_{\tilde{E}23}=M^2_{\tilde{E}32},\nonumber\\
&&T_{e12}=T_{e21},~T_{e13}=T_{e31},~T_{e23}=T_{e32},~\tan\beta.
\end{eqnarray}
Without special statement, the non-diagonal elements of the parameters are supposed as zero.

\subsection{ Muon anomalous magnetic moment}

\begin{figure}[ht]
\setlength{\unitlength}{5mm}
\centering
\includegraphics[width=3.0in]{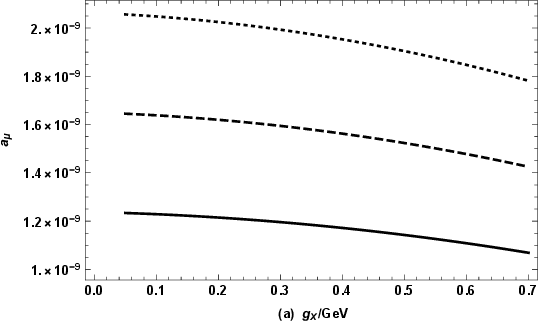}
\includegraphics[width=3.0in]{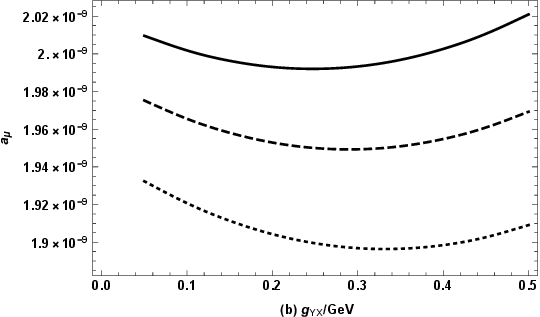}
\vspace{0.2cm}
\caption{$a_\mu$ versus $g_X$(a) and $g_{YX}$(b). The dotted line, dashed line and solid line in Fig.\ref{WW}(a) correspond to $\tan\beta$ equal to 50, 40 and 30 respectively. In Fig.\ref{WW}(b), the solid (dashed, dotted) line corresponds to the results with $g_X=0.3~(0.4,~0.5)$. }{\label {WW}}
\end{figure}

In this subsection, we study the one-loop $g-2$ in $U(1)_X$SSM model by MIA and expect to get some inspiration about using MIA to find LFV. The new experiment data of muon $g-2$ is reported by the workers at Fermilab National Accelerator Laboratory (FNAL)\cite{ZZZ1,ZZZ2,ZZZ3,ZZZ4}. Combined with the previous Brookhaven National Laboratory (BNL) E821 result\cite{ZZZ5}, we get the new averaged experiment value of muon anomaly is $a^{\exp}_{\mu}=116592061(41)\times 10^{-11}$(0.35ppm).
 The departure from the SM prediction is $\Delta a_\mu=a^{\exp}_\mu-a^{SM}_\mu=251(59)\times 10^{-11}$, which is about 4.2$\sigma$. We set the particle masses as: $M_1=800~\rm GeV$,~$\mu=350~{\rm GeV}$,~$m_{\tilde{\nu}^R_L}=150~\rm GeV$,~ $m_{\tilde{\nu}^I_L}=140~\rm GeV$,~$m_{\tilde{L}_L}=800~\rm GeV$,~$m_{\tilde{L}_R}=850~\rm GeV$ and $m_{\lambda_{\tilde{X}}}=450~\rm GeV$ to get Fig.\ref{WW}. When $g_{YX}=0.2$ in Fig.\ref{WW}(a), we can see that from bottom to top are solid line~($\tan\beta=30$), dashed line~($\tan\beta=40$) and dotted line~($\tan\beta=50$) and the overall trend of the three lines is downward. That is to say, $\tan\beta$ is a sensitive parameter and larger $\tan\beta$ leads to larger $a_\mu$.

We set $\tan\beta=50$ in Fig.\ref{WW}(b). It is obvious that three lines with the same tendency to decrease and then increase. The dotted line~($g_X=0.5$) is below the dashed line~($g_X=0.4$), and
the dashed line is below the solid line~($g_X=0.3$). When $g_{X}$ increases, $a_\mu$ decreases. The influence of $g_{YX}$ on $a_\mu$ mainly depends on its own value: when $g_{YX}$ is less than 0.3, $g_{YX}$ increases and $a_\mu$ decreases, but when $g_{YX}>0.3$, the situation is just the opposite. The value of $a_\mu$ can reach $2\times10^{-9}$ at most, it can reach about $80\%$ of the departure($\Delta a_\mu$), which better meets the experimental limitations.
The above conclusion is the same as Eq.(\ref{amuS}), so we can find other sensitive parameters more intuitively through formula Eq.(\ref{amuS}).

\subsection{ The processes of $\mu\rightarrow{e\gamma}$}

In order to study the parameters affecting LFV, we need to study some sensitive parameters. To show the numerical results clearly,
we draw the relation diagrams and scatter diagrams of $Br(\mu\rightarrow{e\gamma})$
with different parameters.

\begin{figure}[ht]
\setlength{\unitlength}{5mm}
\centering
\includegraphics[width=3.0in]{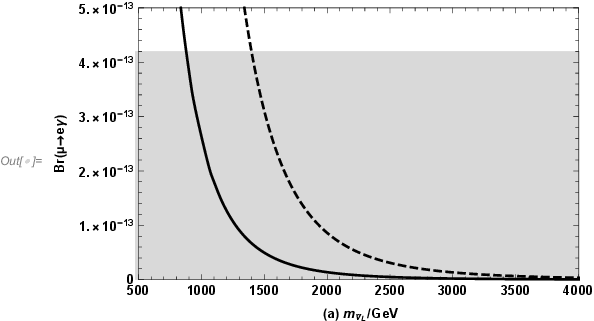}
\vspace{0.2cm}
\setlength{\unitlength}{5mm}
\centering
\includegraphics[width=3.0in]{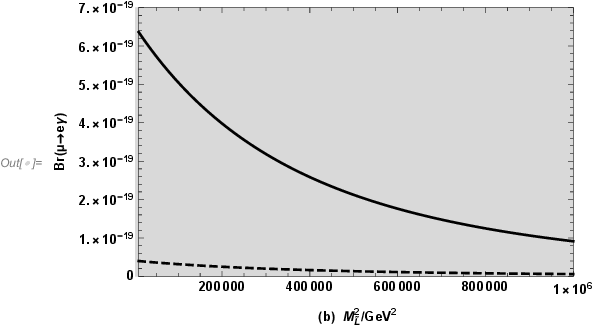}
\vspace{0.2cm}
\setlength{\unitlength}{5mm}
\centering
\includegraphics[width=3.0in]{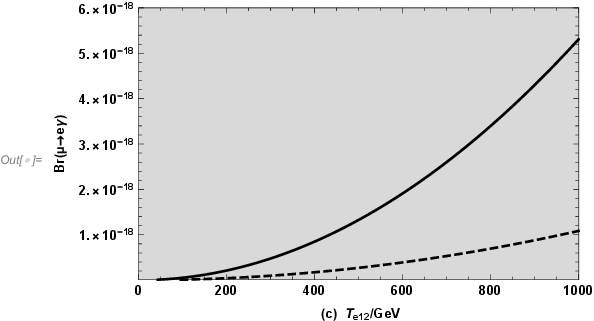}
\caption{$Br(\mu\rightarrow{e\gamma})$ schematic diagrams affected by different parameters. The gray area is reasonable value range, where $Br(\mu\rightarrow{e\gamma})$ is lower than the upper limit.
As $T_{e12}=0$, the dashed and solid lines in Fig.\ref {2}(a) correspond to $M_{\tilde{L}12}^2=500~\rm GeV^2$ and $M_{\tilde{L}12}^2=200~\rm GeV^2$.
The dashed line and solid line respectively represent $T_{e12}=50$ GeV and 100 GeV in Fig.\ref {2}(b).
We set $M_{\tilde{L}}^2=5\times10^5~\rm TeV^2$, the dashed line($\tan\beta=9$) and solid line($\tan\beta=20$) in Fig.\ref {2}(c) are generated.}{\label {2}}
\end{figure}

  The gray area is current limit on LFV decay $\mu\rightarrow{e\gamma}$ in Fig.\ref{2}. With the parameters $M_{\tilde{L}12}^2=0$,~$M_{\tilde{E}12}^2=0$ and $T_{e12}=0$, we plot $Br(\mu\rightarrow{e\gamma})$
  versus $m_{\tilde{\nu}_L}$ in the Fig.\ref {2}(a). The dashed curve corresponds to $M^2_{\tilde{L}12}=500~\rm GeV^2$ and
  the solid line corresponds to $M^2_{\tilde{L}12}=200~\rm GeV^2$. On the whole, both lines show a downward trend. $m_{\tilde{\nu}_L}$ and $Br(\mu\rightarrow{e\gamma})$ are inversely proportional. The smaller $m_{\tilde{\nu}_L}$ is, the greater the value of $Br(\mu\rightarrow{e\gamma})$ is. Separately, the dashed line is larger than the solid line, and the ranges consistent with the experimental value are $1400~\rm GeV-4000~\rm GeV$ and $900~\rm GeV-1400~\rm GeV$ respectively. $m^2_{\tilde{L}12}$ and $Br(\mu\rightarrow{e\gamma})$ are positively correlated. If the value of $M^2_{\tilde{L}12}$ gets smaller, the value of $m_{\tilde{\nu}_L}$ can be less than 1000 GeV.

  We show $Br(\mu\rightarrow{e\gamma})$ varying with $M^2_{\tilde{L}}$ by the solid curve ($T_{e12}=100$ GeV) and dashed curve ($T_{e12}=50$ GeV) in the Fig.\ref {2}(b). We can see that the overall values meet the limit, and the trend is a subtractive function, and the solid line is greater than the dotted line. So we can conclude that as $T_{e12}$ increases, $Br(\mu\rightarrow{e\gamma})$ also increases. When $M^2_{\tilde{L}}$ increases, $Br(\mu\rightarrow{e\gamma})$ decreases. The numerical results are tiny and at the order of $10^{-19}$.

Finally, we analyze the effects of the parameter $T_{e12}$ on branching ratio of $\mu\rightarrow{e\gamma}$. The numerical results are shown in the Fig.\ref {2}(c) by the dashed curve ($\tan\beta=9$) and solid curve ($\tan\beta=20$). The value of the solid line is greater than that of the dashed line, and both show an upward trend. Therefore, The relationship between $\tan\beta$ and $Br(\mu\rightarrow{e\gamma})$, $T_{e12}$ and $Br(\mu\rightarrow{e\gamma})$ is the similar, and they are all positively correlated.

\begin{figure}[h]
\setlength{\unitlength}{5mm}
\centering
\includegraphics[width=3.0in]{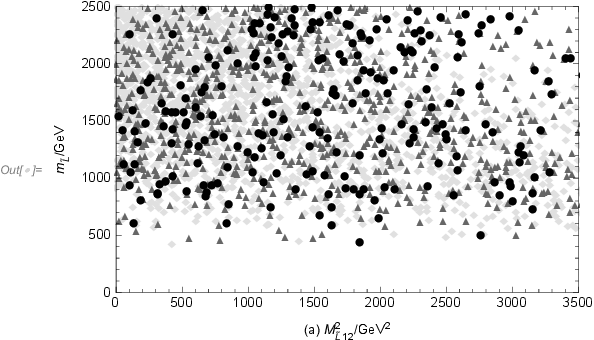}
\vspace{0.2cm}
\setlength{\unitlength}{5mm}
\centering
\includegraphics[width=3.0in]{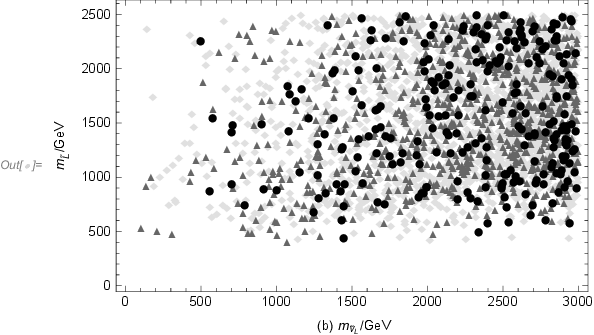}
\vspace{0.2cm}
\setlength{\unitlength}{5mm}
\centering
\includegraphics[width=3.0in]{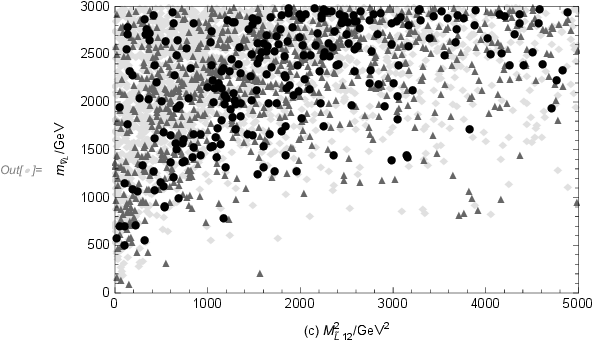}
\vspace{0.2cm}
\setlength{\unitlength}{5mm}
\centering
\includegraphics[width=3.0in]{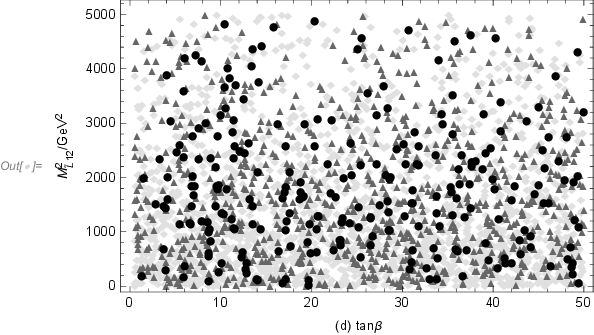}
\caption{Under the premise of lower current limit on lepton flavor violating decay $\mu\rightarrow e\gamma$, reasonable parameter space is selected to scatter points, where \textcolor{light-gray} {$\blacklozenge$} mean the value of $Br(\mu\rightarrow e\gamma)$ less than $1.5\times10^{-13}$, \textcolor{dark-gray}{$\blacktriangle$} mean $Br(\mu\rightarrow e\gamma)$ in the range of $1.5\times10^{-13}$ to $3.5\times10^{-13}$, $\bullet$ show $3.5\times10^{-13}$ $\leq Br(\mu\rightarrow e\gamma)<4.2\times10^{-13}$.}
{\label {3}}
\end{figure}

For more multidimensional analysis of sensitive parameters, we scatter points according to Table \ref{III}~(Part of $\mu\rightarrow{e\gamma}$) to get Fig.\ref{3}.
We set the range of  \textcolor{light-gray} {$\blacklozenge$} (0$<Br(\mu\rightarrow e\gamma)<1.5\times10^{-13}$), \textcolor{dark-gray}{$\blacktriangle$} ($1.5\times10^{-13}\leq Br(\mu\rightarrow e\gamma)<3.5\times10^{-13}$) and $\bullet$ ($3.5\times10^{-13}$ $\leq Br(\mu\rightarrow e\gamma)<4.2\times10^{-13}$) to
represent the results in different parameter spaces for the process of $\mu\rightarrow{e\gamma}$.

 The relationship between $M^2_{{\tilde{L}12}}$ and $m_{\tilde{L}}$ is shown in Fig.\ref {3}(a).
\textcolor{light-gray} {$\blacklozenge$} are mainly concentrated in the upper left corner, then the outer layer are \textcolor{dark-gray}{$\blacktriangle$} and finally $\bullet$.  When $M^2_{{\tilde{L}12}}$ is near 0 and $m_{\tilde{L}}$ is near $2500 ~\rm GeV$, $Br(\mu\rightarrow e\gamma)$ gets the minimum value.
Fig.\ref {3}(b) is plotted in the plane of $m_{\tilde{L}}$ versus $m_{\tilde{\nu}_L}$. We can clearly find that the points are mainly concentrated in the right, and the color gradually deepens from lower left to upper right.
The effects of $M_{\tilde{L}12}^2$ and $m_{\tilde{\nu}_L}$ on $Br(\mu\rightarrow e\gamma)$ are shown in the Fig.\ref {3}(c).
All points are mainly concentrated in the upper left corner and on both sides of axis $M_{\tilde{L}12}^2=0$ and axis $m_{\tilde{\nu}_L}=3000~\rm GeV$. From the inside to the outside, they are \textcolor{light-gray} {$\blacklozenge$},\textcolor{dark-gray}{$\blacktriangle$} and $\bullet$. When $M_{\tilde{L}12}^2$ becomes larger and $m_{\tilde{\nu}_L}$ becomes smaller, the value of $Br(\mu\rightarrow e\gamma)$ increases to reach the experimental measurement.
Fig.\ref {3}(d) shows the effect of $\tan\beta$ and $M_{\tilde{L}12}^2$ on $Br(\mu\rightarrow e\gamma)$. All points are mainly concentrated near the x-axis, and the value increases from bottom to top.

\begin{table*}
\begin{tabular*}{\textwidth}{@{\extracolsep{\fill}}|l|llll@{}}
\hline
\diagbox{Parameters}{Range}{processes}&\makecell[c]{$\mu\rightarrow{e\gamma}$\\$(j=2,i=1)$}&\makecell[c]{$\tau\rightarrow{\mu\gamma}$\\$(j=3,i=2)$}&\makecell[c]{$\tau\rightarrow{e\gamma}$\\$(j=3,i=1)$}\\
\hline
$\tan\beta$&$0.5\sim50$&$0.5\sim50$&$0.5\sim50$\\
$M^2_{\tilde{L}ij}/\rm GeV^2$&$0\sim5000$&$0\sim5000$&$0\sim5000$\\
$M^2_{\tilde{E}ij}/\rm GeV^2$&$0\sim10^4$&$0\sim10^4$&$0\sim10^4$\\
$T_{eij}/\rm GeV$&$-1\sim1$&$-50\sim50$&$-50\sim50$\\
$m_{\tilde{\nu}_L}/\rm GeV$&$100\sim3000$&$100\sim3000$&$100\sim3000$\\
$m_{\tilde{L}}/\rm GeV$&$400\sim2500$&$400\sim2500$&$400\sim2500$\\
\hline
\end{tabular*}
\caption{Scanning parameters for Fig.{\ref {3},\ref {5},\ref {6}}. Without special statement, the non-zero values of non-diagonal elements $m^2_{\tilde{L}ij},m^2_{\tilde{E}ij},T_{eij}$ corresponding to $l_j\rightarrow{l_i\gamma}$ are shown in the column.}\label{III}
\end{table*}

\subsection{ The processes of $\tau\rightarrow{\mu\gamma}$}

\begin{figure}[h]
\setlength{\unitlength}{5mm}
\centering
\includegraphics[width=2.8in]{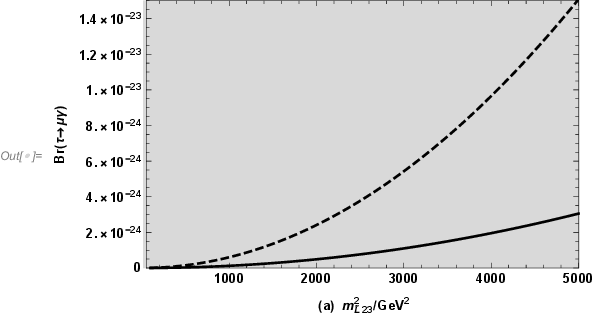}
\vspace{0.2cm}
\setlength{\unitlength}{5mm}
\centering
\includegraphics[width=2.8in]{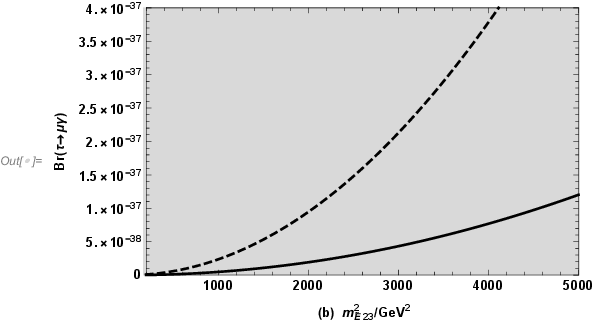}
\caption{Below the experimental limit, the line diagram of parameters and $Br(\tau\rightarrow\mu\gamma)$.
In Figs.\ref {4}(a)(b), solid lines and dotted lines represent $\tan\beta=9$ and $\tan\beta=20$.}{\label{4}}
\end{figure}

To study the influence of parameters $M^2_{\tilde{L}23}$ and $M^2_{\tilde{E}23}$ on $Br(\tau\rightarrow{\mu\gamma})$ in Fig.\ref {4},
we suppose the parameters  $M_{\tilde{L}12}^2=0$,~$M_{\tilde{E}12}^2=0$ and $T_{e12}=0$  and plot the solid line ($\tan\beta=9$) and dashed line ($\tan\beta=20$). In Fig.\ref {4}(a), we can see that $M^2_{\tilde{L}23}$ corresponds to $Br(\tau\rightarrow{\mu\gamma})$. We plot $M^2_{\tilde{E}23}$ varying with $Br(\tau\rightarrow{\mu\gamma})$ in the Fig.\ref {4}(b). Both figures show an upward trend within the experimental limit, and the dashed line is larger than the solid line, so we can draw a conclusion: when $M^2_{\tilde{L}23}$ or $M^2_{\tilde{E}23}$ increases, $Br(\tau\rightarrow{\mu\gamma})$ also increases. In the whole, the numerical results in Fig.\ref {4} are very tiny.

\begin{figure}[h]
\setlength{\unitlength}{5mm}
\centering
\includegraphics[width=2.8in]{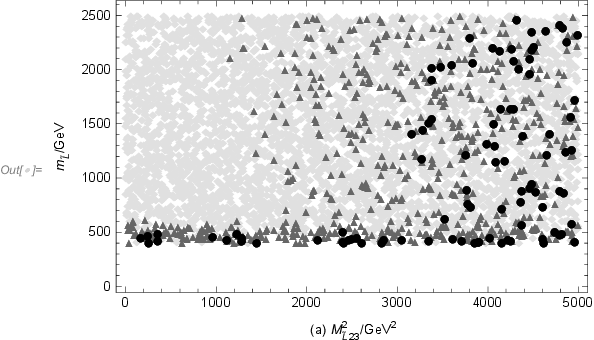}
\setlength{\unitlength}{5mm}
\centering
\includegraphics[width=3.0in]{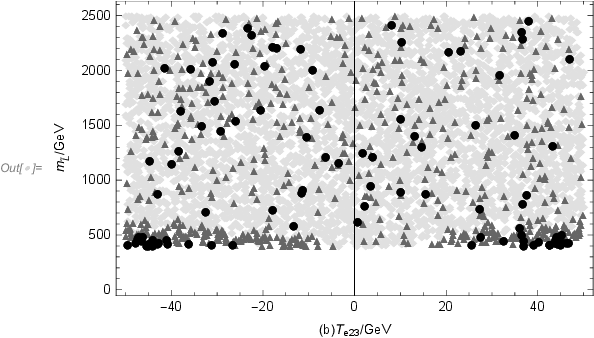}
\vspace{0.2cm}
\setlength{\unitlength}{5mm}
\centering
\includegraphics[width=3.0in]{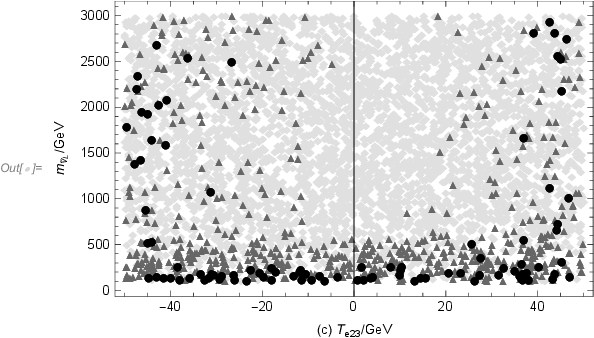}
\vspace{0.2cm}
\setlength{\unitlength}{5mm}
\centering
\includegraphics[width=3.0in]{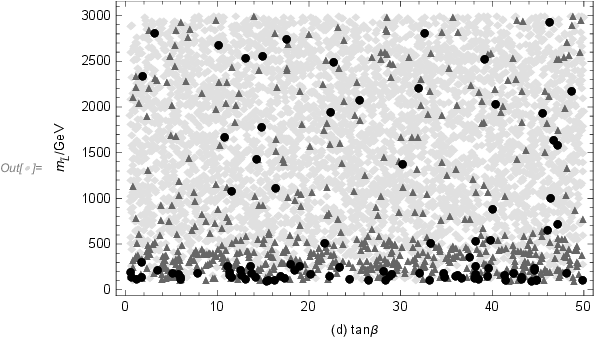}
\caption{ Scatter points under the restriction of the upper limit of $Br(\tau\rightarrow\mu\gamma)$. \textcolor{light-gray} {$\blacklozenge$} represent 0$<Br(\tau\rightarrow\mu\gamma)<1\times10^{-10}$, \textcolor{dark-gray}{$\blacktriangle$} represent $1\times10^{-10}\leq Br(\tau\rightarrow\mu\gamma)<9\times10^{-10}$ and $9\times10^{-10}\leq Br(\tau\rightarrow\mu\gamma)<4.4\times10^{-8}$ are represented by $\bullet$.}{\label {5}}
\end{figure}

We scatter points according to the parameters given in Table \ref {III}~(part of $\tau\rightarrow{\mu\gamma}$) to obtain Fig.\ref {5}.
Where \textcolor{light-gray} {$\blacklozenge$}, \textcolor{dark-gray}{$\blacktriangle$} and $\bullet$ represent (0$<Br(\tau\rightarrow{\mu\gamma})<1.0\times10^{-10}$),  ($1.0\times10^{-10}$ $\leq Br(\tau\rightarrow{\mu\gamma})<9.0\times10^{-10}$) and  ($9.0\times10^{-10}$$ \leq Br(\tau\rightarrow{\mu\gamma})<4.4\times10^{-8}$) respectively.

 $M_{\tilde{L}23}^2$ corresponds to $m_{\tilde{L}}$ in Fig.\ref {5}(a). Horizontally, \textcolor{light-gray} {$\blacklozenge$} are mainly concentrated in $0<M_{\tilde{L}23}^2<1500~\rm GeV^2$, \textcolor{dark-gray}{$\blacktriangle$} are in $1500~\rm GeV^2$$<M_{\tilde{L}23}^2<3200~\rm GeV^2$, and $\bullet$  are distributed in $3200~\rm GeV^2$$<M_{\tilde{L}23}^2<5000~\rm GeV^2$. Vertically, \textcolor{dark-gray}{$\blacktriangle$} and $\bullet$ are concentrated near $m_{\tilde{L}}=500~\rm GeV$, and there are obvious stratification, from bottom to top are $\bullet$, \textcolor{dark-gray}{$\blacktriangle$}, \textcolor{light-gray} {$\blacklozenge$}. So we can know that as $M_{\tilde{L}23}^2$ increases, $Br(\tau\rightarrow{\mu\gamma})$ increases, and when $m_{\tilde{L}}$ increases, $Br(\tau\rightarrow{\mu\gamma})$ decreases.
We plot $m_{\tilde{L}}$ varying with $T_{e23}$ in the Fig.\ref {5}(b). The three types of points are almost symmetrical about $T_{e23}=0$. The smaller the $m_{\tilde{L}}$ is, the greater the value of $Br(\tau\rightarrow{\mu\gamma})$. The farther away the value of $T_{e23}$ from the 0 axis, the greater the value of $Br(\tau\rightarrow{\mu\gamma})$.
Fig.\ref {5}(c) is shown in the plane of $T_{e23}$ versus $m_{\tilde{\nu}_L}$, where
the centralized distributions of the three types of points are distributed in a "U" shape on both sides of the $T_{e23}=0$ axis. \textcolor{light-gray} {$\blacklozenge$} distribute on the innermost side, followed by \textcolor{dark-gray}{$\blacktriangle$} and $\bullet$ on the outermost side. So as $m_{\tilde{\nu}_L}$ increases, $Br(\tau\rightarrow{\mu\gamma})$ decreases.
Finally, we analyze the effects from parameters $\tan\beta$ and $m_{\tilde{L}}$ in Fig.\ref {5}(d). All points are \textcolor{light-gray} {$\blacklozenge$}, \textcolor{dark-gray}{$\blacktriangle$} and $\bullet$ from top to bottom. The smaller the value of $m_{\tilde{L}}$, the larger the value of $Br(\tau\rightarrow{\mu\gamma})$.

\subsection{ The processes of $\tau\rightarrow{e\gamma}$}
\begin{figure}[h]
\setlength{\unitlength}{5mm}
\centering
\includegraphics[width=3.0in]{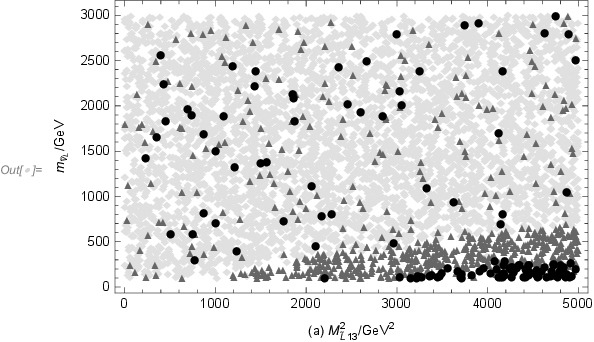}
\vspace{0.2cm}
\setlength{\unitlength}{5mm}
\centering
\includegraphics[width=3.0in]{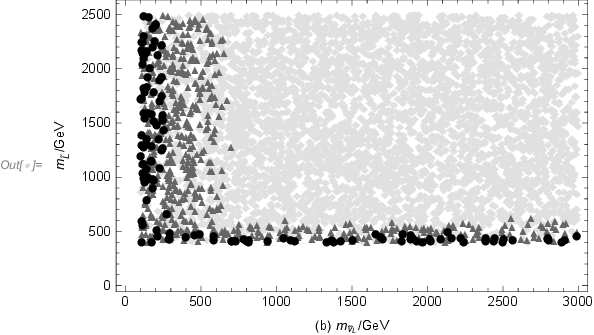}
\vspace{0.2cm}
\setlength{\unitlength}{5mm}
\centering
\includegraphics[width=3.0in]{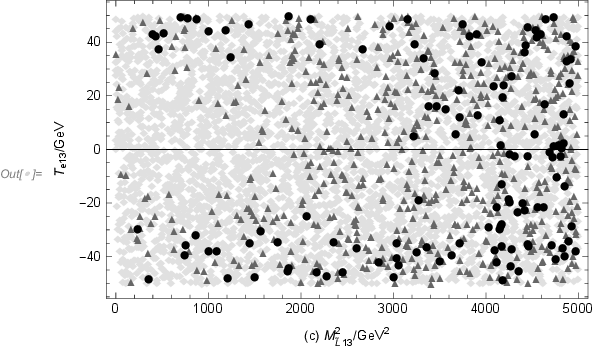}
\caption{{\label{6}}  For the scatter diagrams of the parameters below the experimental limit $Br(\tau\rightarrow{e\gamma})$, different points represent the different ranges of $Br(\tau\rightarrow{e\gamma})$. \textcolor{light-gray} {$\blacklozenge$} represent less than $1.0\times10^{-10}$. \textcolor{dark-gray}{$\blacktriangle$} represent the range of $1.0\times10^{-10}$ to $8.0\times10^{-10}$, and $\bullet$ represent the range of $8.0\times10^{-10}$ to $3.3\times10^{-8}$.}
\end{figure}

 Based on the Table \ref{III}~(part of $\tau\rightarrow{e\gamma}$), we analyze $\tau\rightarrow{e\gamma}$ to study the possibility of LFV in Fig.\ref {6}.
 The branching ratio of $\tau\rightarrow{e\gamma}$ process is denoted by: \textcolor{light-gray} {$\blacklozenge$} (0$<Br(\tau\rightarrow{e\gamma})<1.0\times10^{-10}$), \textcolor{dark-gray}{$\blacktriangle$} ($1\times10^{-10}$ $\leq Br(\tau\rightarrow{e\gamma})<8.0\times10^{-10}$) and $\bullet$ ($Br(\tau\rightarrow{e\gamma})$ from $8.0\times10^{-10}$ to $3.3\times10^{-8}$).

   The Fig.\ref {6}(a) shows the effects from $m_{\tilde{\nu}_L}$ and $M_{\tilde{L}13}^2$.
   Most points are concentrated in lower right quarter. $\bullet$ are on the innermost side of the whole region.
   \textcolor{dark-gray}{$\blacktriangle$} are in the middle and  \textcolor{light-gray} {$\blacklozenge$} are on the outermost side. The numerical performance is that the larger the $M_{\tilde{L}13}^2$ and the smaller the $m_{\tilde{\nu}_L}$, the larger the $Br(\tau\rightarrow{e\gamma})$.
 In Fig.\ref {6}(b), we analyze the effects of $m_{\tilde{\nu}_L}$ and $m_{\tilde{L}}$ on $Br(\tau\rightarrow{e\gamma})$. In the whole figure, $\bullet$ are mainly close to both sides of the x-axis and y-axis, and then \textcolor{dark-gray}{$\blacktriangle$} with the same trend, and the rest is \textcolor{light-gray} {$\blacklozenge$}. $Br(\tau\rightarrow{e\gamma})$ decreases with the increase of ~$m_{\tilde{\nu}_L}$ and $m_{\tilde{L}}$.
  Fig.\ref {6}(c) has two axes $m_{\tilde{L}13}^2$ versus $T_{e13}$. All three points show "$\supset$" shaped distribution, from left to right are \textcolor{light-gray} {$\blacklozenge$},  \textcolor{dark-gray}{$\blacktriangle$}, $\bullet$. So $Br(\tau\rightarrow{e\gamma})$ increases with the increase of $T_{e13}$.

\section{discussion and conclusion}

The $U(1)_X$SSM has new superfields including righ-handed neutrinos and three Higgs superfields $\hat{\eta},~\hat{\bar{\eta}},~\hat{S}$, and its local
gauge group is $SU(3)_C\times SU(2)_L \times U(1)_Y\times U(1)_X$.
We use MIA to study muon anomalous magnetic moment. Combined with the latest experimental data, our numerical results can reach about $2\times10^{-9}$, which can better fit the measurement result, and play a certain role in promoting the study of LFV.
We use the method of MIA to study lepton flavor violating decays $l_j\rightarrow{l_i\gamma}$ in the $U(1)_X$SSM model.
From the order of magnitude of branching ratio and data analysis, we can find that the restriction on lepton flavor violation in the process of $\mu\rightarrow{e\gamma}$ is stronger. This provides a reference for other lepton flavor violation work in the future.

  We take into account the constraints from the upper limits on LFV branching ratios of
$l_j\rightarrow l_i\gamma$. In the numerical calculation, we take many parameters as variables including $\tan\beta,~g_X,~g_{YX},~M_{\tilde{L}}^2,~M_{\tilde{L}ij}^2,~M_{\tilde{E}}^2,~M_{\tilde{E}ij}^2,~ \delta^{AB}_{ij},~m_{\tilde{L}},~m_{\tilde{\nu}_L}$ and $T_{eij}$. Through the analysis of the numerical results, we find that $M_{\tilde{L}ij}^2,~M_{\tilde{E}ij}^2,~g_{YX},~\delta^{AB}_{ij},~m_{\tilde{L}},~m_{\tilde{\nu}_L}$ and $T_{eij}$ are sensitive parameters. $Br(l_j\rightarrow{l_i\gamma})$ is an increasing function of $M_{\tilde{L}ij}^2,~M_{\tilde{E}ij}^2,~T_{eij},~g_{YX},~\delta^{AB}_{ij}$, and decreasing function of $m_{\tilde{L}}$ and $m_{\tilde{\nu}_L}$.  $ g_X$ can also give influence on the numerical results but not very large. That is to say they give mild influences on the numerical results. Finally, we come to the conclusion that the non-diagonal elements which correspond to the generations of the initial lepton and final lepton
are main sensitive parameters and LFV sources.

{\bf Acknowledgments}

This work is supported by National Natural Science Foundation of China (NNSFC)
	(No. 11535002, No. 11705045), Natural Science Foundation of Hebei Province
	(A2020201002). Post-graduate's Innovation Fund Project of Hebei University
    (HBU2022ss028).


\begin{thebibliography}{50}

\vspace{3mm}
\bibitem{1}
K. Abe et al. (T2K Collaboration), Phys. Rev. Lett. {\bf 107} (2011)
041801; J. Ahn et al. (RENO Collaboration), Phys.
Rev. Lett. {\bf 108} (2012) 191802; F. An et al. (DAYABAY
Collaboration), Phys. Rev. Lett. {\bf108} (2012) 171803.




\bibitem{2}S. T. Petcov, Sov. J. Nucl. Phys. {\bf25} (1977) 340 JINR-E2-10176.
\bibitem{02}
K.~S.~Sun, J.~B.~Chen and X.~Y.~Yang, et al.,
Chin. Phys. C \textbf{43} (2019) 043101.



\bibitem{3} U. Ellwanger, C. Hugonie, and A.M. Teixeira, Phys. Rep. {\bf496} (2010) 1-77.


\bibitem{UU1}
B.~Yan, S.~M.~Zhao and T.~F.~Feng,
Nucl. Phys. B \textbf{975} (2022) 115671 .

\bibitem{Sarah1}	F. Staub,  [arXiv: 0806.0538].
\bibitem{Sarah2}	 F. Staub, Comput. Phys. Commun. {\bf185} (2014) 1773.
\bibitem{Sarah3}	F. Staub, Adv. High Energy Phys. {\bf2015} (2015) 840780.
\bibitem{MSSM} J. Rosiek, Phys. Rev. D {\bf41} (1990) 3464.
\bibitem{sm1} CMS collaboration, Phys. Lett. B {\bf716} (2012) 30; ATLAS collaboration, Phys. Lett. B {\bf716} (2012) 1.
\bibitem{220725}  V. Cirigliano, K. Fuyuto, C. Lee, et al., JHEP {\bf03} (2021) 256.
\bibitem{9}  K. S. Sun , T. Guo and W. Li, et al., Eur. Phys. J. C {\bf80} (2020)  1167.

\bibitem{21}
S.~M.~Zhao, T.~F.~Feng and H.~B.~Zhang, et al., Phys. Rev. D \textbf{92} (2015) 115016.

\bibitem{new1}
T.~Nomura, H.~Okada and Y.~Uesaka,
Nucl. Phys. B \textbf{962} (2021) 115236.


\bibitem{new2}
A.~Ilakovac, A.~Pilaftsis and L.~Popov,
Phys. Rev. D \textbf{87} (2013) 053014.

\bibitem{20}
T.~T.~Wang, S.~M.~Zhao and X.~X.~Dong, et al., JHEP \textbf{04} (2022) 122.

\bibitem{04}
S.~M.~Zhao, L.~H.~Su, and X.~X.~Dong, et al.,
JHEP \textbf{03}  (2022) 101.

\bibitem{07}
E.~Arganda, M.~J.~Herrero, and R.~Morales, et al.,
JHEP \textbf{03} (2016) 055.



\bibitem{06}
E.~Arganda, M.~J.~Herrero and X.~Marcano, et al.,
Phys. Rev. D \textbf{95} (2017) 095029.

\bibitem{05}
M.~J.~Herrero, X.~Marcano and R.~Morales, et al.,
Eur. Phys. J. C \textbf{78}  (2018) 815.



\bibitem{08}
G.~Haghighat and M.~M.~Najafabadi,
[arXiv:2204.04433 [hep-ph]].

\bibitem{10} Particle Data Group, Prog. Theor. Exp. Phys. {\bf2020} (2020) 083C01.

\bibitem{UU3}S.M. Zhao, T.F. Feng and  M.~J.~Zhang,  et al., JHEP {\bf02} (2020) 130.

\bibitem{LCTHiggs1} M. Carena, J.R. Espinosaos and C.E.M. Wagner, et al., Phys. Lett. B {\bf 355} (1995) 209.
\bibitem{LCTHiggs2} M. Carena, S. Gori and N.R. Shah, et al., JHEP {\bf 1203} (2012) 014.
\bibitem{UMSSM5}G. Belanger, J.D. Silva and H.M. Tran, Phys. Rev. D {\bf95} (2017) 115017.

\bibitem{B-L1}V. Barger, P.F. Perez and S. Spinner, Phys. Rev. Lett. {\bf 102} (2009) 181802.
 	
\bibitem{B-L2}
P.H. Chankowski, S. Pokorski and J. Wagner, Eur. Phys. J. C {\bf 47} (2006) 187.

\bibitem{gaugemass}J.L. Yang, T.F. Feng and S.M. Zhao, et al., Eur. Phys. J. C {\bf78} (2018) 714.




\bibitem{09}
T.~Moroi,
Phys. Rev. D \textbf{53} (1996) 6565-6575.


 \bibitem{su1}CMS Collaboration, Phys. Lett. B {\bf716} (2012) 30.
 \bibitem{su2}A TLAS Collaboration, Phys. Lett. B {\bf716} (2012) 1.



\bibitem{xin1}G.~Aad  et al. [ATLAS],
Phys. Lett. B \textbf{796} (2019) 68-87.

\bibitem{ZPG1} G. Cacciapaglia, C. Csaki, G. Marandella, et al.,
Phys. Rev. D {\bf74} (2006) 033011.

\bibitem{ZPG2} M. Carena, A. Daleo and  B. A. Dobrescu, et al., Phys. Rev. D {\bf70} (2004) 093009.

\bibitem{TanBP} L. Basso, Adv. High Energy Phys. {\bf2015} (2015) 980687.


\bibitem{A2021}
P.~Athron, C.~Bal\'azs and D.~H.~J.~Jacob, et al., JHEP \textbf{09} (2021) 080.



\bibitem{ZZZ1}
T.~Albahri, et al., [Muon g-2],
Phys. Rev. D \textbf{103} (2021)  072002.


\bibitem{ZZZ2}
M.~Endo, K.~Hamaguchi and S.~Iwamoto, et al.,
JHEP \textbf{07} (2021) 075.



\bibitem{ZZZ3}
M.~Chakraborti, L.~Roszkowski and S.~Trojanowski,
JHEP \textbf{05} (2021) 252.


\bibitem{ZZZ4}
F.~Wang, L.~Wu and Y.~Xiao, et al.,
Nucl. Phys. B \textbf{970} (2021) 115486.




\bibitem{ZZZ5}
G.~W.~Bennett, et al., [Muon g-2],
Phys. Rev. D \textbf{73} (2006) 072003.

\end{thebibliography}
\end{document}